\documentclass[a4paper,11pt]{article}
\usepackage{siunitx}
\usepackage{jheppub}
\usepackage{array}
\usepackage{float}
\usepackage{booktabs}
\usepackage{makecell}
\usepackage{caption}
\usepackage{amsmath}
\usepackage{graphicx,tabularx}

\usepackage{gensymb}
\usepackage{url}
\usepackage{footmisc}
\usepackage{amsfonts}
\usepackage{cancel}
\usepackage{color}
\usepackage{multirow} 
\usepackage{amssymb}
\usepackage{pifont}
\usepackage{epstopdf}
\usepackage{slashed}
\usepackage{comment}
\usepackage{booktabs} 
\usepackage{mathrsfs}
\usepackage{subcaption}
\usepackage[toc,page]{appendix}
\usepackage{mathtools}
\usepackage{romannum}
\usepackage[normalem]{ulem}
\usepackage{bbold}
\usepackage{xcolor}
\usepackage{tikz-feynman}
\usepackage{tikz}
\usepackage{placeins}
\usepackage{hyperref}
\usepackage{orcidlink}
\usepackage[normalem]{ulem}
\allowdisplaybreaks

\usepackage{lineno}
\begin{document}
\begin{flushright}
\preprint{MI-HET-882}
\end{flushright}
\title{\boldmath Analytical and Machine Learning Methods for \\
Model Discernment at CE$\nu$NS Experiments}

\author[a]{Iain A. Bisset,}
\author[a]{Bhaskar Dutta,}
\author[b]{Doojin Kim,}
\author[c]{Samiran Sinha,}
\author[d]{Joel W. Walker}

\affiliation[a]{Mitchell Institute for Fundamental Physics and Astronomy, Department of Physics and Astronomy, Texas A$\&$M University, College Station, TX 77843, USA}

\affiliation[b]{Department of Physics, University of South Dakota, Vermillion, SD 57069, USA}

\affiliation[c]{Department of Statistics, Texas A$\&$M University, College Station, TX 77843, USA}

\affiliation[d]{Department of Physics and Astronomy, Sam Houston State University, Huntsville, TX 77341, USA}

\emailAdd{ibisset@tamu.edu}
\emailAdd{dutta@physics.tamu.edu}
\emailAdd{doojin.kim@usd.edu}
\emailAdd{sinha@stat.tamu.edu}
\emailAdd{jwalker@shsu.edu}

\abstract{
Neutrino experiments are often limited by low statistics, sizable systematic uncertainties, and coarse observable binning, which can hinder discrimination among competing beyond-the-Standard-Model (BSM) explanations of anomalous signals. In particular, analyses based primarily on total event-rate differences are vulnerable to source-normalization uncertainties and to degeneracies among models that induce similar inclusive yields. Using stopped-pion coherent elastic neutrino-nucleus scattering (CE$\nu$NS) as a benchmark environment, we study how much model-discrimination power can be obtained from correlations in baseline, recoil energy, and timing that are less sensitive to the total rate. As benchmark BSM scenarios, we consider a $3+1$ sterile-neutrino framework and neutral-current non-standard neutrino interactions (NSI). We show with a likelihood-based analysis that these scenarios can be distinguished in nontrivial regions of parameter space once multidimensional shape information is retained. We further demonstrate with convolutional neural networks that substantial discrimination remains possible even after the total event rate is explicitly removed from the input, indicating that the relevant information is genuinely encoded in the shape of the CE$\nu$NS distribution. Finally, through multi-class classification within the sterile parameter space, we show that in favorable regions the same observables can support approximate localization of the underlying sterile-neutrino benchmark point. Our results highlight the complementary roles of conventional and machine-learning-based inference in moving neutrino new-physics searches from anomaly detection to physics interpretation.
}

\maketitle

\section{Introduction}
Predictions of the neutrino-sector of the Standard Model (SM) have been confronted with an increasingly broad program of precision measurements in oscillation, scattering, and source-based experiments. As the number of precision tests in the neutrino sector continues to grow, it is becoming increasingly important to develop robust methods for model selection and parameter estimation tailored to the challenges inherent in these experiments. In particular, the wide landscape of viable new-physics scenarios makes effective model discrimination an increasingly important task. New-physics signals associated with neutrinos across a wide range of experiments can arise from diverse mechanisms. These include, for example, oscillation-based scenarios involving sterile neutrinos~\cite{Acero_2024,Bisset2024} as well as non-oscillation explanations such as non-standard neutrino interactions~\cite{Proceedings:2019qno} or other modifications of neutrino production, propagation, and detection. In general, such scenarios can alter the observed data in several distinct ways: through a change in the overall event yield, through distortions in spectral, spatial, and/or temporal distributions, and through modified flavor structure. In practice, however, neutrino experiments often operate in regimes where limited statistics, imperfect flux knowledge, detector effects, and/or coarse binning reduce the extent to which these signatures can be cleanly disentangled. As a result, different new-physics models may generate qualitatively similar excesses or deficits at the level of inclusive rates, even when their underlying dynamical origins are very different.

This general issue is especially relevant for stopped-pion neutrino experiments. In these experiments, a high intensity proton beam with energy $\sim 1$ GeV hits a fixed target and the detectors are kept at a distance $\sim$ 20 meters from the target. In many analyses of such facilities, sensitivity has traditionally been driven by simple event selections together with comparisons of total or partially integrated event rates~\cite{COHERENT:2017ipa, COHERENT:2020iec, COHERENT:2021xmm,CCM:2021leg}. More recently, however, both phenomenological studies~\cite{Dutta:2019nbn, Dutta:2020vop, Bisset2024} and experimental analyses~\cite{COHERENT:2021pvd, CCM:2021leg, CCM:2021yzc} have highlighted the usefulness of the intrinsic time structure of stopped-pion sources for new-physics searches: the prompt $\nu_\mu$ component and the delayed $\nu_e/\bar{\nu}_\mu$ components populate different timing windows, so that timing information can be used to perform a partial flavor decomposition and thereby sharpen sensitivity to nonstandard scenarios. In this sense, the separation into prompt and delayed samples may already be viewed as a coarse form of shape-based analysis, built from correlations among the available observables. This observation motivates the present study, in which we ask how much further discriminatory power can be gained by treating baseline, recoil energy, and timing together within a more systematic multidimensional inference framework.

For this reason, the question of discovery is only part of a broader scientific problem. Once one goes beyond inclusive event counts and begins to exploit shape information encoded in the available observables, a more refined set of questions becomes accessible.
If an anomalous excess or deficit is observed in a neutrino experiment, it is essential to determine not only whether the data prefer a nonstandard interpretation over the SM, but also which model class is favored and which region of parameter space is selected by the data. Such information is critical for connecting the observed anomaly to other measurements. A preferred parameter region inferred from one experiment can imply correlated signals, null tests, or exclusion targets in independent experiments, thus sharpening the program of confirmation and accelerating progress toward a consistent physical interpretation.

Apart from the motivation provided by possible nonstandard signals at future experiments, several past experiments have also exhibited anomalous results that can accommodate a range of different model-dependent interpretations, further necessitating the development of new model discernment methodologies. While many observations are well described within the standard three-neutrino framework, several longstanding anomalies and tensions continue to motivate the exploration of beyond-the-SM (BSM) physics. Representative examples arise from short-baseline, pion decay at rest, and reactor experiments. 

Notable anomalies include the LSND excess~\citep{LSND:2001aii} and the MiniBooNE low-energy excess~\citep{MiniBooNE:2008yuf,MiniBooNE:2018esg,MiniBooNE:2020pnu}.
Results from SBL experiments, such as BEST~\cite{Barinov:2022wfh}, when combined with SAGE~\cite{SAGE:1998fvr} and GALLEX~\cite{GALLEX:1997lja}, provide evidence consistent with a $3+1$ sterile neutrino framework. 
However, the $3+1$ scenario is not preferred by the recent results from MicroBooNE~\cite{MicroBooNE:2025nll}. 
Although the interpretation of these results remains unsettled, they provide concrete motivation for developing analysis strategies capable not only of identifying departures from standard expectations but also of determining which classes of new-physics scenarios may account for an observed signal.

Recent years have also seen the widespread application of machine-learning (ML) techniques to neutrino physics~\cite{Backhouse:2015xva, Aurisano:2016jvx, MicroBooNE:2016dpb, MicroBooNE:2018kka, Baldi:2018qhe, KM3NeT:2020zod, DUNE:2020gpm, MicroBooNE:2020hho, MicroBooNE:2020yze, Abbasi_2024, Villarreal2026}. In many contexts, ML has demonstrated excellent performance in event classification, background rejection, and high-dimensional feature extraction (see e.g.,~\cite{Psihas:2020pby, Kim:2021pcz, Franceschini:2022vck}). This naturally raises an important question for neutrino-sector new-physics searches: beyond improving signal sensitivity or detecting anomalous behaviors, can ML methods also enhance our ability to identify the underlying model responsible for a deviation from the SM, and to infer the corresponding parameter point with meaningful precision? Addressing this question requires going beyond a binary signal-versus-background framework and instead treating the problem as one of structured inference in a multidimensional observable space. Related ML approaches have also recently been explored for model discrimination in collider settings, including at the LHC~\cite{Agrawal:2026lvg,Moreno:2026mqk}.

In this work, we investigate these issues in the context of sterile-neutrino-induced and/or nonstandard neutrino interaction-induced modifications to coherent elastic neutrino-nucleus scattering (CE$\nu$NS) at stopped-pion neutrino sources.  This environment is particularly well-suited to the present study. Because the neutrino production mechanism is time structured (charged-pion-induced prompt neutrino signals vs. muon-induced delayed ones), and because detectors can in principle be deployed at multiple baselines, the signal is encoded not only in the total event rate but also in the correlated shapes of the recoil-energy, timing, and baseline distributions. These observables provide a richer arena for model discrimination than rate-only measurements and allow one to test how effectively different analysis strategies extract nonstandard information from realistic datasets.

Our study has two primary goals. First, we compare a traditional likelihood-based analysis with a ML-based analysis in their ability to distinguish sterile-neutrino scenarios from alternative nonstandard interaction (NSI) scenarios. This comparison allows us to assess model discrimination power using both conventional and high-dimensional methods within a common benchmark framework. Second, within the sterile-neutrino hypothesis itself for illustration, we study how well one can identify the underlying parameter point from the observed CE$\nu$NS data, including a multi-class classification strategy for approximate localization within the sterile parameter space. In other words, we examine not only whether a nonstandard signal can be detected, but also how accurately the relevant model parameters can be reconstructed once such a signal is present.

The broader aim is to clarify the role that modern inference tools can play in moving neutrino new-physics searches from anomaly detection to physics interpretation and, in favorable cases, toward approximate parameter-space localization. If successful, such methods can provide a pathway from an observed deviation, to a favored model hypothesis, to concrete predictions for independent experimental confirmation. All of the techniques that we are developing can also be applied to distinguish light dark matter scenarios from those involving NSI and sterile neutrinos discussed so far. 

The rest of this paper is structured as follows. In Sec.~\ref{sec:benchmarks}, we introduce the benchmark physics setup used throughout this work, including CE$\nu$NS at stopped-pion sources, the benchmark sterile-neutrino and NSI scenarios, and the statistical framework for constructing the binned event distributions. In Sec.~\ref{Analytical}, we present a conventional likelihood-based analysis and quantify the resulting model-discrimination reach in the benchmark parameter space. In Sec.~\ref{Architecture}, we describe the common neural-network architecture and training procedure adopted for the ML studies. In Sec.~\ref{sec:MLdisc}, we report the results of two complementary ML tasks: binary classification for model discrimination between the sterile-neutrino and best-fit NSI hypotheses, and multi-class classification for identifying the underlying region of sterile-neutrino parameter space. Finally, in Sec.~\ref{sec:conlcusions}, we summarize our main findings and discuss their implications for future neutrino and dark-sector searches.

\section{Benchmark Physics Cases} \label{sec:benchmarks}

Before outlining the analysis framework and presenting our main results, we first specify the benchmark scenarios that will serve as the basis of our study. As briefly discussed in the Introduction, we focus on neutrino-induced CE$\nu$NS signals at stopped-pion neutrino experiments  and consider the possibility that observable deviations from the standard three-neutrino prediction originate from nonstandard neutrino physics. As representative examples, we study the $3+1$ sterile-neutrino oscillation scenario and scenarios with NSIs.

\subsection{CE$\nu$NS at stopped-pion sources}
To study model discrimination and parameter inference in a controlled setting, we consider benchmark CE$\nu$NS signals induced by neutrinos from stopped-pion sources. Specifically, we envision a (hypothetical) experimental configuration consisting of multiple CsI detectors placed at different baselines from the neutrino source. This setup is motivated by existing and proposed stopped-pion neutrino programs, such as COHERENT at Oak Ridge National Laboratory~\citep{COHERENT:2021xmm}, Coherent CAPTAIN-Mills (CCM) at Los Alamos National Laboratory~\citep{CCM:2021leg}, and the Search for Hidden Neutrinos at the European Spallation Source~\cite{Soleti:2023hlr}. In addition to its previous runs with CsI and LAr, the COHERENT collaboration is currently operating, or plans to operate NaI, Ge, Neon, CsI, and LAr detectors, all located at distances of approximately $19-28$ meters~\cite{COHERENT:2026ewu}. The CCM experiment is likewise planning to deploy multi-ton segmented CsI detectors in addition to its ongoing LAr detector~\cite{CCM_CSI}. Our purpose here is not to model any one experiment in complete detail, but rather to define a realistic benchmark environment in which baseline, recoil-energy, and timing information can all be exploited simultaneously.

In stopped-pion neutrino experiments, intense GeV-scale proton beams strike a target and copiously produce charged pions. The $\pi^-$ rapidly lose energy and are typically captured before decaying, while the $\pi^+$ come to rest and decay via $\pi^+ \to \mu^+ \nu_\mu$. The decay product $\mu^+$ subsequently also stops and decays through $\mu^+ \to e^+ \nu_e \bar{\nu}_\mu$, thereby yielding the characteristic neutrino flux of stopped-pion sources. The two-body and three-body nature of $\pi^+$ and $\mu^+$ decays, respectively, implies that the $\nu_\mu$ spectrum is mono-energetic while the $\bar{\nu}_\mu$ and $\nu_e$ spectra span a continuous range of energies. Specifically, the differential fluxes for each source component are given as follows:
\begin{eqnarray}
    \label{prompt-energy}
    \frac{d\Phi_{\nu_{\mu}}}{dE_{\nu_{\mu}}} &=& \mathcal{N} \delta \left(E_{\nu_{\mu}} - \frac{m_{\pi}^{2} -m_{\mu}^{2}}{2m_{\pi}} \right), \\
    \label{delayed-energy-mubar}
    \frac{d\Phi_{\bar{\nu}_{\mu}}}{dE_{\bar{\nu}_{\mu}}} &=& \mathcal{N} \frac{16}{m_{\mu}^{4}} \left( 3m_{\mu}E_{\bar{\nu}_{\mu}}^2 - 4E_{\bar{\nu}_{\mu}}^3  \right), \\
    \label{delayed-energy-e}
    \frac{d\Phi_{\nu_{e}}}{dE_{\nu_{e}}} &=& \mathcal{N} \frac{96}{m_{\mu}^{4}} \left( m_{\mu}E_{\nu_{e}}^2 - 2E_{\nu_{e}}^3  \right), 
\end{eqnarray}
where $\mathcal{N}$ is the normalization parameter given by $\mathcal{N}=\dfrac{rN_{\rm POT}}{4\pi L^2}$ with $r$, $N_{\rm POT}$, and $L$ denoting the pion production fraction, the number of protons on target (POT), and the source-to-detector distance, respectively. For our benchmark scenario, we take $r = 0.05$ and $N_{\rm POT} = 10^{22}$. The baseline $L$ varies for each detector in the experiment. We assume a setup consisting of several equal-mass 50kg CsI detectors logarithmically distributed in length between 5m and 25m from the neutrino source.

Equally important for our purposes is the time structure of the neutrino source. Because the pion lifetime ($\sim 26$~ns in the rest frame) is much shorter than the muon lifetime ($\sim 2.2~\mu$s in the rest frame), the prompt $\nu_{\mu}$ component is temporally separated from the delayed $\nu_{e}$ and $\bar{\nu}_{\mu}$ components. We model the timing distributions as
\begin{eqnarray}
    \label{prompt-time}
    \frac{d\Phi_{\nu_{\mu}}}{dt} &=& \frac{1}{b\tau_{\pi}\sqrt{2\pi}} \int_{0}^{t} \exp \left(-\frac{(t' - a)^2}{2b^2} \right) \exp \left( -\frac{(t - t')}{\tau_{\pi}} \right) dt'\,, \\
    \label{delayed-time}
    \frac{d\Phi_{(\bar{\nu}_{\mu}, \nu_{e})}}{dt} &=& \frac{1}{\tau_{\mu}} \int_{0}^{t} \frac{d\Phi_{\nu_{\mu}}}{dt'} \exp \left( -\frac{(t - t')}{\tau_{\mu}} \right) dt'\,,
\end{eqnarray}
where $\tau_\pi$ and $\tau_\mu$ denote the pion and muon lifetimes, while $a$ and $b$ parameterize the proton-pulse time profile. In our benchmark setup, these timing parameters are taken to be the values adopted in the COHERENT analysis~\cite{COHERENT:2021xmm}.

Since both $\pi^+$ and $\mu^+$ decay at rest, the resulting neutrinos are emitted isotropically.  They reach the CsI detector of interest, and scatter off the cesium and iodine nuclei primarily via the CE$\nu$NS process, which involves the neutral current and is flavor-agnostic.~\cite{Freedman:1973yd,Scholberg:2005qs}. The differential cross section with respect to the nuclear recoil energy $E_r$ is given by~\cite{Freedman:1973yd,Scholberg:2005qs,Formaggio:2012cpf} 
\begin{equation}
    \label{cevns-cross}
    \frac{d\sigma}{dE_r} = \frac{G_{F}^{2}Q_{V}^{2}M}{2\pi} F^{2}(q^{2}) \left[ 1 - \left( \frac{M E_r}{E_{\nu}^{2}} \right ) + \left( 1 - \frac{E_r}{E_{\nu}} \right)^{2} \right],
\end{equation}
where $M$ is the total mass of the target nucleus, $E_{\nu}$ is the incoming neutrino energy, and $F(q^2)$ parametrizes the nuclear form factor that encodes the loss of coherence at finite momentum transfer $q$, for which the Helm form factor~\cite{Helm:1956zz} is commonly adopted. The effective SM vector charge is $Q_{V} = (2g_{u} + g_{d})Z + (g_{u} + 2g_{d})N$, where $g_{u}$ and $g_{d}$ are the vector couplings to the $u$- and $d$-type quarks, and $Z$ and $N$ are respectively the proton and neutron numbers of the nucleus under consideration.

A key advantage of the CE$\nu$NS process, and the detectors designed to measure it (for example, this also includes sapphire, silicon, and germanium crystal detectors with superconducting transition edge sensors \cite{Mirzakhani:2025bqz,Iyer:2020nxe}), is their ability to measure low-energy recoils while maintaining a significant detector mass.  This is crucial for taking advantage of the cross-section boost in the low-energy regime, which also scales quadratically (proportional to the neutron count squared) with the size of the target nucleus.  However, one downside of such detectors is that they are unable to directly recover the incident neutrino energy, because the scattering angle is not measured.  As such, if a neutrino source presents a continuous spectrum, there will be smearing of the oscillation response.  However, this can be mitigated by binning in the recoil energy, and more energetic neutrinos are still likely to produce harder scatters, on average~\cite{Dutta:2015nlo}.  This also highlights the advantage of binning in time, since timing information can help disentangle the discrete (monochromatic) and continuous spectral components of the neutrino source. In particular, the oscillation response of the monochromatic component does not suffer from the smearing described above.

The key point is that this framework provides three complementary observables: recoil-energy spectra, timing distributions, and variation with detectors placed across multiple length baselines. These observables respond differently to changes in neutrino flavor content, oscillation parameters, and interaction structure. Consequently, even when different new-physics scenarios generate similar modifications to the total event rate, they may still be distinguishable once the full multidimensional information is incorporated. This is precisely the feature that makes stopped-pion CE$\nu$NS a useful benchmark for the likelihood-based and machine-learning-based analyses presented in the following sections.

\subsection{Benchmark BSM scenarios} \label{sec:benchmarkmodels}

To study model discrimination and parameter reconstruction in a controlled setting, we consider benchmark datasets generated under the SM and two representative BSM scenarios.

\subsubsection{Benchmark I: $3+1$ sterile-neutrino oscillation scenario}

As our first benchmark new-physics framework, we consider the $3+1$ sterile-neutrino scenario. In this extension of the standard three-neutrino picture, the active neutrino sector is supplemented by one additional, predominantly sterile, mass eigenstate $\nu_4$. The corresponding leptonic mixing matrix is then enlarged from a $3\times 3$ matrix to a $4\times 4$ matrix. At short baselines, where the oscillation effects driven by $\Delta m_{21}^2$ and $\Delta m_{31}^2$ can be neglected, the phenomenology is controlled primarily by the larger mass-squared splitting
\begin{equation}
\Delta m_{41}^2 \equiv m_4^2 - m_1^2,
\end{equation}
together with the mixing matrix elements $U_{\alpha 4}$ involving the fourth mass eigenstate, where $\alpha=e,\mu,\tau$.

In the short-baseline limit, the oscillation probabilities take a two-flavor-like form. For appearance channels ($\alpha\neq\beta$), one has
\begin{equation}
P_{\alpha\beta}(E_\nu,L)
=
\sin^2 2\theta_{\alpha\beta}\,
\sin^2\!\left(\frac{\Delta m_{41}^2 L}{4E_\nu}\right),
\qquad (\alpha\neq\beta),
\label{sterile-Prob-app}
\end{equation}
whereas for disappearance channels,
\begin{equation}
P_{\alpha\alpha}(E_\nu,L)
=
1-\sin^2 2\theta_{\alpha\alpha}\,
\sin^2\!\left(\frac{\Delta m_{41}^2 L}{4E_\nu}\right).
\label{sterile-Prob-dis}
\end{equation}
The effective mixing amplitudes are given by
\begin{equation}
\sin^2 2\theta_{\alpha\beta}
=
4|U_{\alpha 4}|^2 |U_{\beta 4}|^2,
\qquad (\alpha\neq\beta),
\label{angles-app}
\end{equation}
and
\begin{equation}
\sin^2 2\theta_{\alpha\alpha}
=
4|U_{\alpha 4}|^2\left(1-|U_{\alpha 4}|^2\right).
\label{angles-dis}
\end{equation}

In the stopped-pion CE$\nu$NS setup considered here, the neutrino source consists of prompt $\nu_\mu$ and delayed $\nu_e$ and $\bar\nu_\mu$ components. Since CE$\nu$NS is a neutral-current process and is therefore not directly sensitive to the flavor identity among active neutrinos, the dominant observable effect of sterile mixing arises from the depletion of the active neutrino flux through oscillations into the sterile state. Consequently, the relevant parameters for our analysis are primarily $\Delta m_{41}^2$, $|U_{e4}|^2$, and $|U_{\mu4}|^2$. For simplicity, and because the source does not contain an initial $\nu_\tau$ component, we set
\(
|U_{\tau 4}|^2 = 0
\)
throughout this study.

This benchmark scenario provides a well-motivated and economical test case for our analysis framework. It generates characteristic distortions in the baseline and energy dependence of the CE$\nu$NS event distribution, while the prompt-delayed timing structure of stopped-pion sources allows the effects associated with the $\nu_\mu$ and $\nu_e/\bar\nu_\mu$ source components to be partially disentangled. These features make the $3+1$ framework a useful benchmark for assessing both model discrimination and parameter reconstruction.

\subsubsection{Benchmark II: neutral-current NSIs}
As the second benchmark new-physics framework, we consider neutral-current NSIs of neutrinos with quarks. In a model-independent low-energy description, such effects can be parameterized by effective four-fermion operators that modify the neutrino-matter interaction strength relative to the SM. The corresponding effective Lagrangian is written as~\cite{Proceedings:2019qno}
\begin{equation}
    \label{NSI_Lagrangian}
    \mathcal{L}_{\rm NSI}
    =
    -2\sqrt{2}\,G_F\,\epsilon_{\alpha\beta}^{fH}
    \left(\bar{\nu}_{\alpha}\gamma_{\sigma}P_L\nu_{\beta}\right)
    \left(\bar{f}\gamma^{\sigma}P_H f\right),
\end{equation}
where $\alpha,\beta=e,\mu,\tau$ label neutrino flavors, $f$ denotes the matter fermion, and $H=L,R$ specifies the chirality of the fermion current. For CE$\nu$NS, the relevant target constituents are the light quarks $u$ and $d$, and only the vector combinations $\epsilon_{\alpha\beta}^{fV} \equiv \epsilon_{\alpha\beta}^{fL} + \epsilon_{\alpha\beta}^{fR}$ contribute coherently at leading order. In this sense, CE$\nu$NS provides a particularly clean probe of neutral-current NSI.

The presence of NSI modifies the effective weak charge of the nucleus and, consequently, the CE$\nu$NS event rate. For an incident neutrino of flavor $\alpha$, the squared effective coupling becomes
\begin{multline}
    \label{NSI_total_coupling}
    \left(Q_{\alpha}^{\rm NSI}\right)^2
    =
    \left(
    \left[
    2\left(g_{u}+\epsilon^{uV}_{\alpha\alpha}\right)
    +
    \left(g_{d}+\epsilon^{dV}_{\alpha\alpha}\right)
    \right] Z
    +
    \left[
    \left(g_{u}+\epsilon^{uV}_{\alpha\alpha}\right)
    +
    2\left(g_{d}+\epsilon^{dV}_{\alpha\alpha}\right)
    \right] N
    \right)^2
    \\
    +
    \sum_{\beta\neq\alpha}
    \left[
    \left(2\epsilon^{uV}_{\alpha\beta}+\epsilon^{dV}_{\alpha\beta}\right)Z
    +
    \left(\epsilon^{uV}_{\alpha\beta}+2\epsilon^{dV}_{\alpha\beta}\right)N
    \right]^2 .
\end{multline}
Here the diagonal couplings $\epsilon^{fV}_{\alpha\alpha}$ shift the flavor-conserving interaction strength, while the off-diagonal couplings $\epsilon^{fV}_{\alpha\beta}$ with $\alpha\neq\beta$ correspond to flavor-changing neutral-current NSI. In both cases, the observable consequence is a modification of the CE$\nu$NS rate and its distribution across the multidimensional observable space used in our analysis.

In contrast to the sterile-neutrino scenario, where the leading CE$\nu$NS signature is primarily a depletion of the active neutrino flux through oscillations into a sterile state, neutral-current NSI directly alter the interaction strength of active neutrinos with the detector material. As a result, NSI can generate distortions that differ in both normalization and flavor dependence. In the stopped-pion CE$\nu$NS environment considered here, this is particularly relevant because the prompt $\nu_\mu$ component and the delayed $\nu_e/\bar{\nu}_\mu$ components populate different regions of the timing distribution. Their contributions can therefore be weighted differently in the presence of flavor-dependent NSI couplings.

This makes neutral-current NSI a useful benchmark complementary to the $3+1$ sterile-neutrino scenario. While both frameworks can induce deviations from the SM expectation in CE$\nu$NS observables, the working mechanisms are physically distinct: sterile mixing modifies the active neutrino flux reaching the detector, whereas NSI modify the scattering response of the detector to the incident active flux. The degree to which these mechanisms can be distinguished forms one of the central questions of the present study.

\subsection{Statistical framework} \label{sec:statframe}

We now specify the statistical framework used to generate and analyze the benchmark event distributions. For a given choice of model parameters---denoted by $\vec{s}$ for the $3+1$ sterile-neutrino scenario and by $\vec{\epsilon}$ for the NSI scenario---the predicted CE$\nu$NS yield is computed in bins of detector baseline, reconstructed recoil energy, and event time. The corresponding expected number of events in each bin is written as
\begin{multline}
     \label{N-Total}
    N_{ijk}^{\textrm{CE$\nu$NS}}\left(\vec{s},\vec{\epsilon}\right)
    =
    \sum_{\alpha}
    \sum_{\beta}
    \sum_{\gamma=\nu_{\mu},\bar{\nu}_{\mu},\nu_{e}}
    \mathcal{N}_{i\beta}
    \int_k d t
    \int_j dE_{\rm rec}
    \int dE_{\rm true}
    \int dE_{\gamma}\;
    \epsilon_T(t)\,\epsilon_E(E_{\rm rec})
    \\
    \times
    P_E(E_{\rm rec},E_{\rm true})\,
    P_{\gamma\alpha}(\vec{s},E_{\gamma},L_i)\,
    \frac{d^2\Phi_{\gamma}}{dE_{\gamma}dt}(E_{\gamma},t,L_i)\,
    \frac{d\sigma_{\alpha\beta}}{dE_{\rm true}}(\vec{\epsilon},E_{\rm true},E_{\gamma}) .
\end{multline}
The labels \{$i$, $j$, $k$\} denote the binning in length, energy, and time, respectively. The numbers of bins in each dimension vary in this analysis, but binning in length is always distributed from 5m to 25m, while binning in recoil energy and timing are always distributed over the ranges defined in the COHERENT CsI experiment~\citep{COHERENT:2021xmm}. The subscripts on the $dt$ and $dE_{\rm rec}$ integrals indicate that the integrations are performed over the time and reconstructed-energy ranges corresponding to the $k$th timing bin and the $j$th $E_{\rm rec}$ bin, respectively.  
The integration over the true nuclear recoil energy, $E_{\rm true}$, goes from 0 to the approximate maximum recoil energy of $\dfrac{2E_{\gamma}^2}
{M}$ for nucleus mass $M$, with $E_{\gamma}$ in the upper bound taken to be $\dfrac{m_\mu}{2}$. The integration over the neutrino energy $E_{\gamma}$ goes from the approximate minimum neutrino energy required to induce a recoil $\sqrt{\dfrac{mE_{\rm true}}{2}}$ to the maximum neutrino energy $\sim\dfrac{m_\mu}{2}$.\footnote{For $\bar{\nu}_\mu$ and $\nu_e$ from the $\mu^+$ decay at rest, the neutrino energy is bounded from above at $\frac{m_\mu^2-m_e^2}{2m_\mu}$, which can be well approximated to $\frac{m_\mu}{2}$ in the $m_e \ll m_\mu$ limit, Using this approximation has no practical impact on our results. For $\nu_\mu$ from the $\pi^+$ decay at rest, the neutrino energy is single-valued at $\frac{m_\pi^2-m_\mu^2}{2m_\pi}$, which lies below this maximum energy. Since $\nu_\mu$ energy distribution is $\delta$-function-like as in Eq.~\eqref{prompt-energy}, taking this integration range does not affect the results in practice.}

The quantity $\mathcal{N}_{i\beta}$ is the number of target nuclei of species $\beta$ in detector $i$, and $L_i$ is the corresponding source-to-detector distance. The factor $\dfrac{d^2\Phi_{\gamma}}{dE_{\gamma}dt}$ represents the differential neutrino flux from the source for initial flavor $\gamma\in\{\nu_\mu,\bar\nu_\mu,\nu_e\}$, including its characteristic energy and timing structure at stopped-pion facilities.

The oscillation factor $P_{\gamma\alpha}(\vec{s},E_{\gamma},L_i)$ gives the probability that a neutrino produced in flavor state $\gamma$ is detected as flavor $\alpha$ after propagating a distance $L_i$. In the sterile-neutrino benchmark, this is the quantity that encodes the dependence on the parameters $\vec{s}$. By contrast, the differential cross section $\dfrac{d\sigma_{\alpha\beta}}{dE_{\rm true}}$ contains the dependence on the NSI parameters $\vec{\epsilon}$ through the modified CE$\nu$NS interaction strength discussed above. The functions $\epsilon_T(t)$ and $\epsilon_E(E_{\rm rec})$ account for the detector efficiencies in timing and reconstructed energy. For simplicity, we simply take $\epsilon_T(t) = 1$ and take $\epsilon_E(E_{\rm rec})$ to be a step function at 6.7 keV. Finally, $P_E(E_{\rm rec},E_{\rm true})$ denotes the detector response function, i.e., the probability density for reconstructing an event with energy $E_{\rm rec}$ given a true recoil energy $E_{\rm true}$. For this response function, we use the formulation given by COHERENT for their CsI CE$\nu$NS detection~\citep{COHERENT:2021xmm}.

Equation~\eqref{N-Total} therefore provides a unified description of the expected multidimensional event distribution in the presence of either sterile-neutrino oscillations or neutral-current NSI. In the SM limit, one simply recovers the corresponding expression by setting $P_{\gamma\alpha}=\delta_{\gamma\alpha}$, the Kronecker delta function, and replacing the cross section with its standard CE$\nu$NS form. This binned event prediction serves as the basic input for both the conventional likelihood analysis and the ML-based inference discussed in the following sections.

\section{Likelihood-Based Model Discrimination}
\label{Analytical}

We first assess the extent to which the benchmark sterile-neutrino and NSI scenarios can be distinguished using a conventional likelihood-based analysis. To this end, we construct a binned likelihood for the CE$\nu$NS event distribution and compare competing hypotheses using the Bayesian information criterion (BIC). This provides a baseline, non-ML approach against which the performance of more flexible inference strategies can later be compared.

\subsection{Likelihood construction and benchmark setup}

Our analysis is based on the binned CE$\nu$NS event prediction introduced in Sec.~\ref{sec:benchmarks}. For a given hypothesis, the expected number of events in bin $(i,j,k)$ is denoted by
\begin{equation}
    \lambda_{ijk} = (1+\eta)\,N_{ijk}^{\textrm{CE$\nu$NS}}(\vec{s},\vec{\epsilon}),
    \label{lambda}
\end{equation}
where $\eta$ is a nuisance parameter that accounts primarily for the overall neutrino-flux normalization uncertainty. The corresponding likelihood for a dataset with observed counts $N^{\rm obs}_{ijk}$ is then written as
\begin{equation}
\mathcal{L}(\vec{s},\vec{\epsilon})
=
\int 
\prod_{(i,j,k)}
\frac{e^{-\lambda_{ijk}}\lambda_{ijk}^{N^{\rm obs}_{ijk}}}{N^{\rm obs}_{ijk}!}
\times \frac{e^{-\eta^2/(2\sigma_\eta^2)}}{\sqrt{2\pi\sigma_\eta^2}}d\eta,
\label{likelihood}
\end{equation}
where the Gaussian prior on $\eta$, which has a mean of zero, has a known standard deviation $\sigma_\eta$ (derived from the experimental setup) and the product is over the total number of bins in length, energy, and time, which we denote as $n_L$, $n_E$, and $n_T$. In what follows, the observed counts are taken from simulated benchmark data generated under a chosen model hypothesis, represented either by $\vec{s}_{\rm choice}$ for the sterile-neutrino scenario or by $\vec{\epsilon}_{\rm choice}$ for the NSI scenario.

To compare competing hypotheses, we use the Bayesian information criterion,
\begin{equation}
    \mathrm{BIC} = k\ln N - 2\ln \mathcal{L}_{\rm max},
    \label{BIC}
\end{equation}
where $k$ is the number of free model parameters, $N = n_L \times n_E \times n_T$ is the total number of bins in the observed dataset, and $\mathcal{L}_{\rm max}$ is the maximum of the likelihood given in (\ref{likelihood}), achieved under the hypothesis being tested. 
The preferred model is identified as the one with the smaller BIC value. For convenience, we also define the corresponding relative likelihood weight
\begin{equation}
    \mathbb{L}(x) = \exp\!\left(\frac{\mathrm{BIC}_{\rm min}-x}{2}\right),
    \label{relative_likelihood}
\end{equation}
where $\mathrm{BIC}_{\rm min}$ denotes the smallest BIC value among the candidate hypotheses for the dataset under consideration, and $x$ is the BIC value of a given hypothesis. This quantity maps the BIC difference with respect to the best-fitting model onto the interval $(0,1]$. In the present analysis, we consider two competing hypotheses, namely the sterile-neutrino and NSI scenarios, such that
\begin{equation}
\mathrm{BIC}_{\rm min}
=
\min\!\left(\mathrm{BIC}_{\rm sterile},\,\mathrm{BIC}_{\rm NSI}\right),
\end{equation}
 within which the SM is a special case 
  with the BSM model parameters being zero. 
Therefore, if a pseudo-dataset is generated under the sterile-neutrino hypothesis, one typically has $\mathrm{BIC}_{\min}=\mathrm{BIC}_{\rm sterile}$, which implies
\(
\mathbb{L}_{\rm sterile}\equiv \mathbb{L}(\mathrm{BIC}_{\rm sterile}) = 1
\)
and
\(
\mathbb{L}_{\rm NSI}\equiv \mathbb{L}(\mathrm{BIC}_{\rm NSI})
= \exp\!\left(-\dfrac{\Delta \mathrm{BIC}}{2}\right)
\)
with
\(
\Delta \mathrm{BIC} \equiv \left|\mathrm{BIC}_{\rm NSI}-\mathrm{BIC}_{\rm sterile} \right|.
\)
For reference, $\Delta \mathrm{BIC}$ values of $2$ and $5$---corresponding to relative likelihoods below $\sim 0.37$ and $\sim 0.08$---are indicative benchmarks for moderate and stronger levels of distinguishability ~\citep{raftery1995bayesian}. 

For the illustrative model-comparison study presented in this section, we restrict attention to two-dimensional parameter subsets of the benchmark new-physics scenarios introduced in Sec.~\ref{sec:benchmarkmodels}. On the sterile-neutrino side, we consider
\begin{equation}
\vec{s}_{\rm choice} = \left\{|U_{e4}|^2,\;\Delta m_{41}^2\right\},
\end{equation}
while on the NSI side we consider
\begin{equation}
\vec{\epsilon}_{\rm choice} = \left\{\epsilon^{uV}_{ee},\;\epsilon^{dV}_{ee}\right\}.
\label{eps}
\end{equation}
Both models therefore have a value of $k = 2$ in Eq. \ref{BIC}. This restricted setup is sufficient to illustrate how the likelihood-based framework responds to two physically distinct mechanisms for generating deviations from the SM expectation. For the purpose of the analyses in this study, we use a sterile-neutrino parameter space window of $10^{-2}-1$ in $|U_{e4}|^2$ and $10^{-1}-10^{3}$ in $\Delta m_{41}^2$. For NSI, we use a window of -1 to 1 in both parameters. 

For each point in the chosen sterile-neutrino parameter window, we generate the corresponding mean CE$\nu$NS dataset, as given in Eq. \ref{lambda} for $\eta = 0$, 
 for a specified binning in baseline, recoil energy, and event time. We then fit the NSI hypothesis to this sterile-generated dataset using \texttt{MultiNest}~\citep{Feroz:2007kg, Feroz:2008xx, Feroz:2013hea}, taking the flux-normalization uncertainty in Eq.~\eqref{likelihood} to be $\sigma_\eta=0.10$. From the selected sterile point and the best NSI hypothesis, we compute the corresponding BIC values and the relative likelihood defined in Eq.~\eqref{relative_likelihood}. By repeating this procedure across the scanned sterile-neutrino parameter space, we obtain a map of the regions in which the sterile benchmark can be distinguished from its best-fitting NSI counterpart for a given experimental configuration and binning choice.

\subsection{Discrimination reach in benchmark parameter space}

\begin{figure*}[t]
\centering
    \subfloat[\{1,1,1\}]{%
        \includegraphics[width=.48\linewidth]{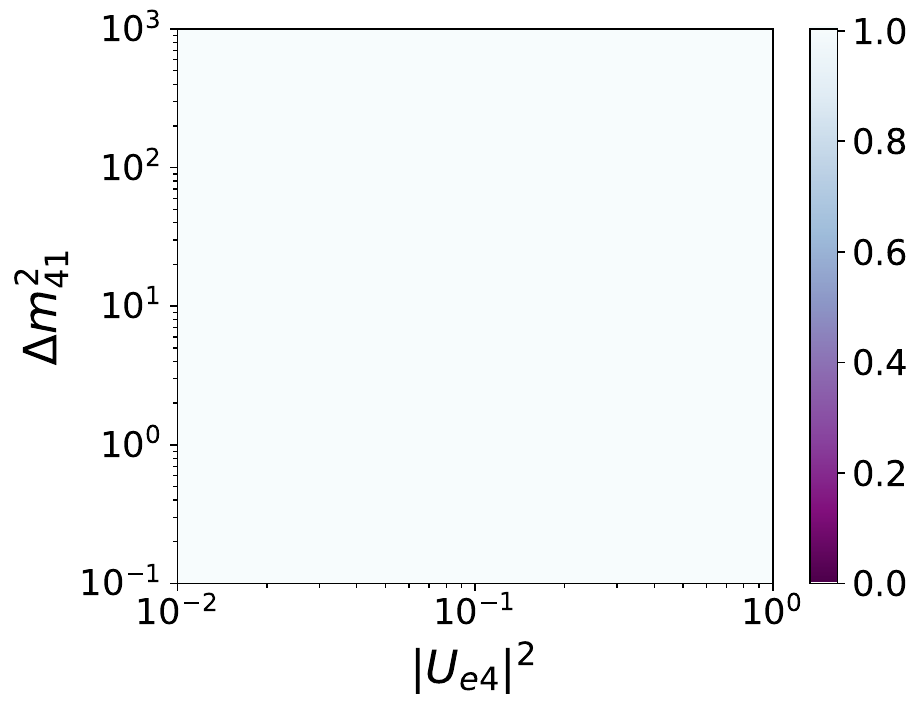}%
        \label{subfig:a}%
    }
    \subfloat[\{1,9,3\}]{%
        \includegraphics[width=.48\linewidth]{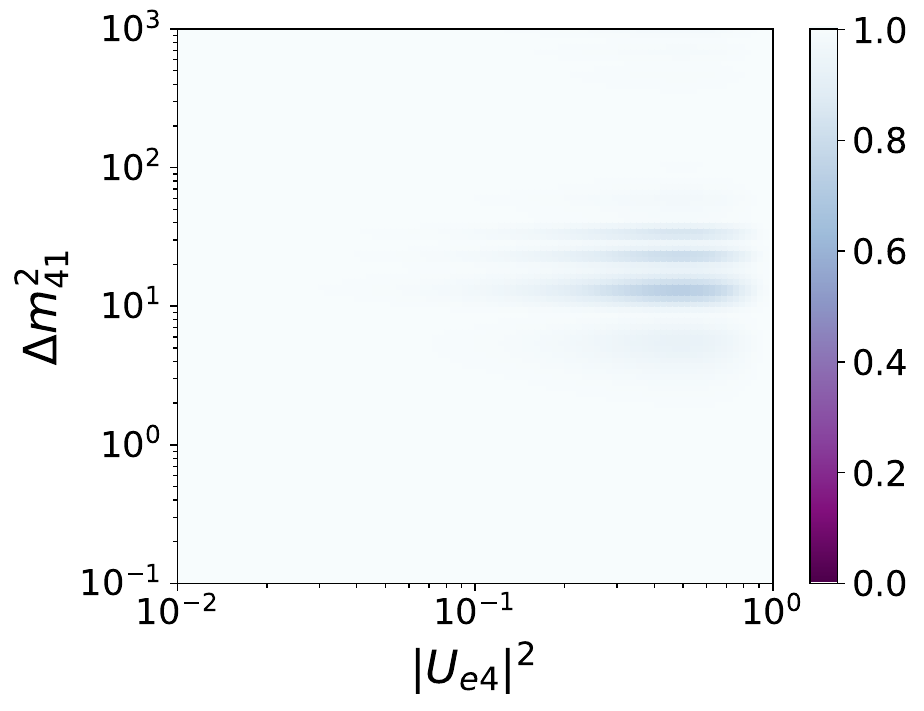}%
        \label{subfig:b}%
    }\\
    \subfloat[\{3,3,1\}]{%
        \includegraphics[width=.48\linewidth]{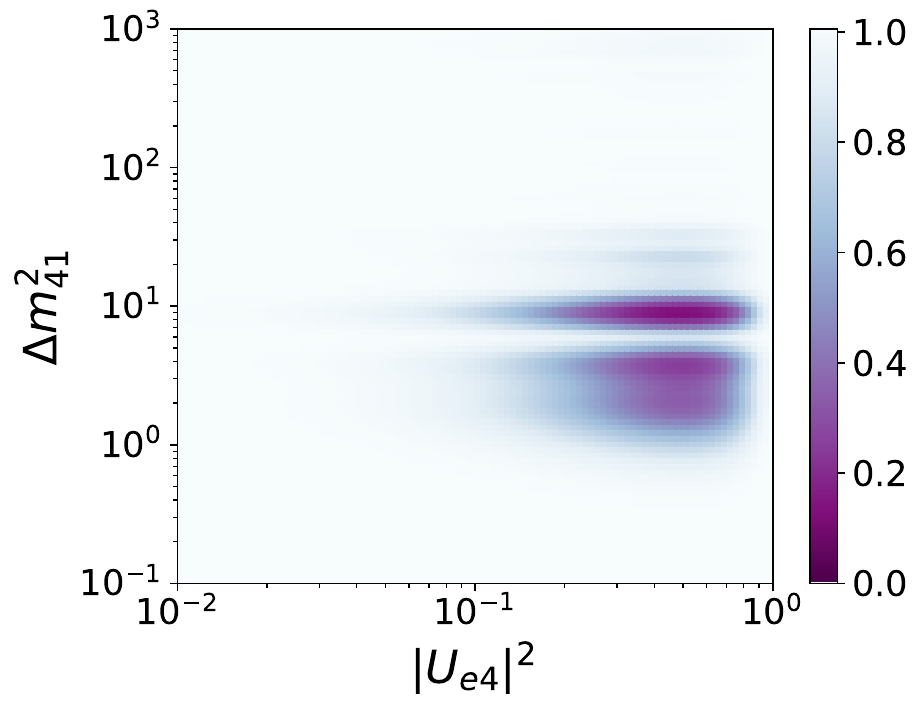}%
        \label{subfig:c}%
    }
    \subfloat[\{3,9,3\}]{%
        \includegraphics[width=.48\linewidth]{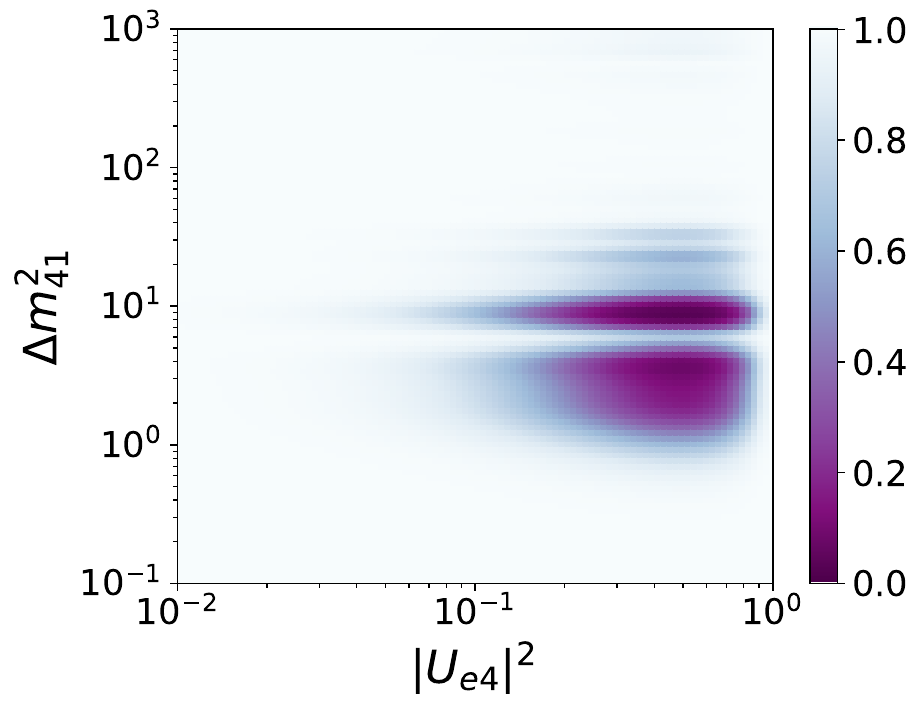}%
        \label{subfig:d}%
    }
    \caption{Discrimination power between the NSI hypothesis and sterile-generated data in the $\nu_e$-coupled sterile benchmark parameter space, shown for several binning configurations. The entries in braces indicate the numbers of bins in \{Length, Energy, Time\}.}
    \label{fig:analytical_1}
\end{figure*}

\begin{figure*}[t]
\centering
    \subfloat[\{9,1,3\}]{%
        \includegraphics[width=.48\linewidth]{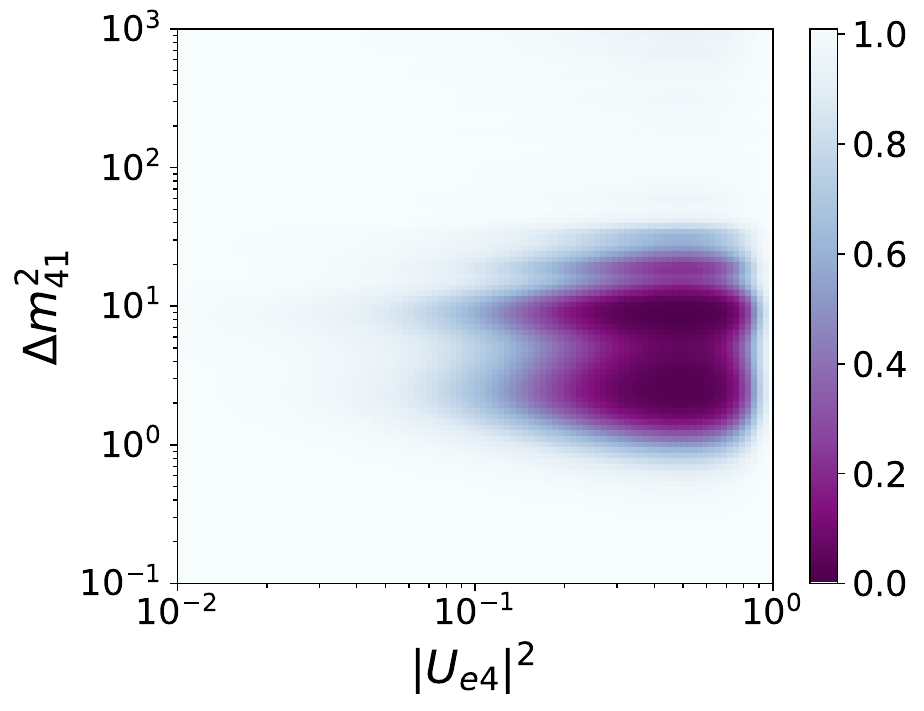}%
        \label{subfig:e}%
    }
    \subfloat[\{9,3,3\}]{%
        \includegraphics[width=.48\linewidth]{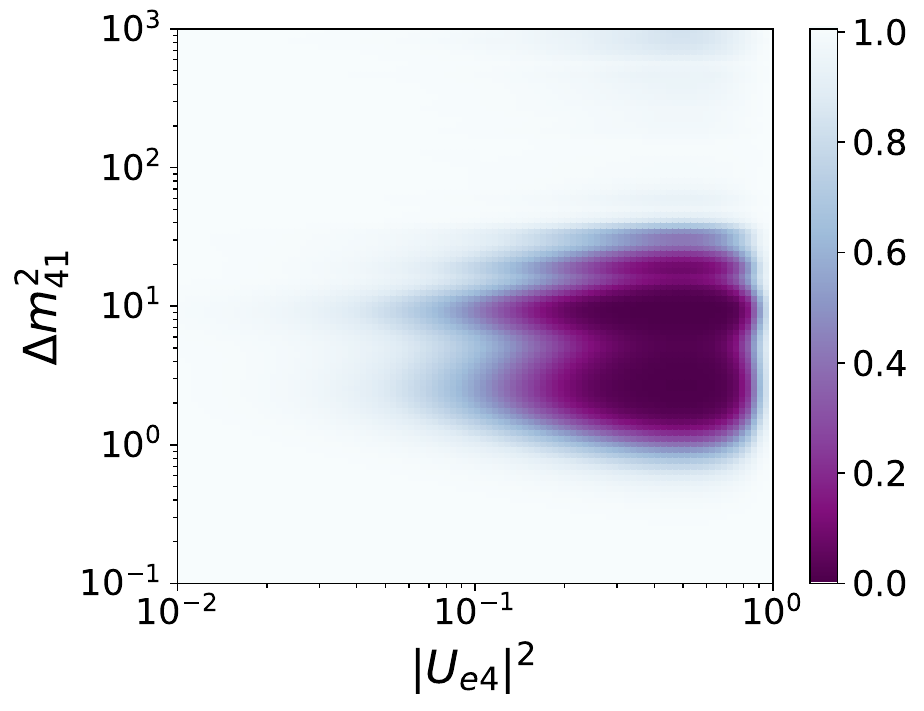}%
        \label{subfig:f}%
    }\\
    \subfloat[\{9,9,1\}]{%
        \includegraphics[width=.48\linewidth]{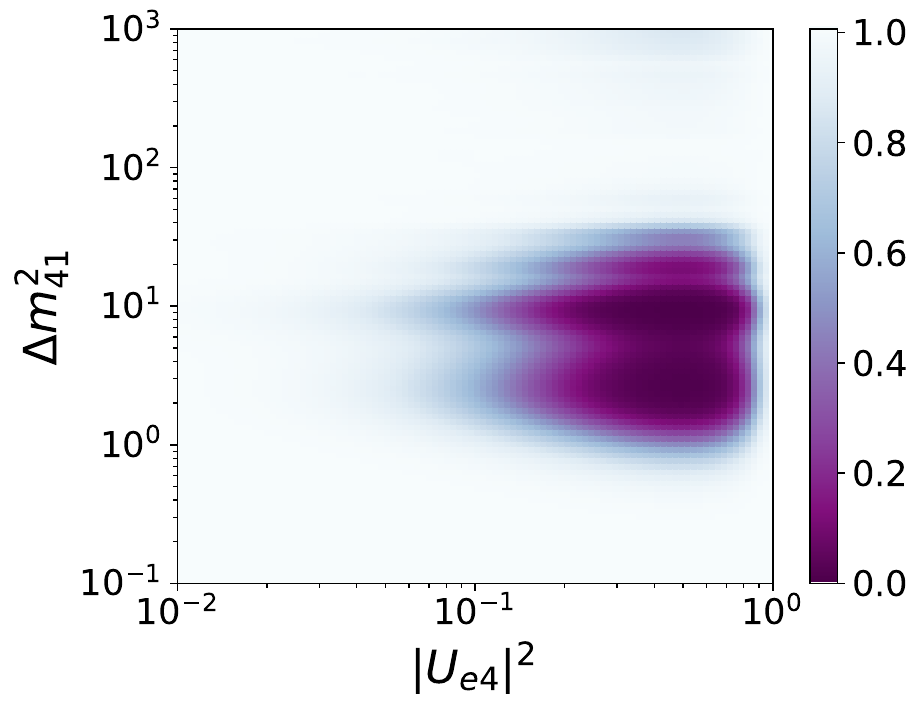}%
        \label{subfig:g}%
    }
    \subfloat[\{9,9,3\}]{%
        \includegraphics[width=.48\linewidth]{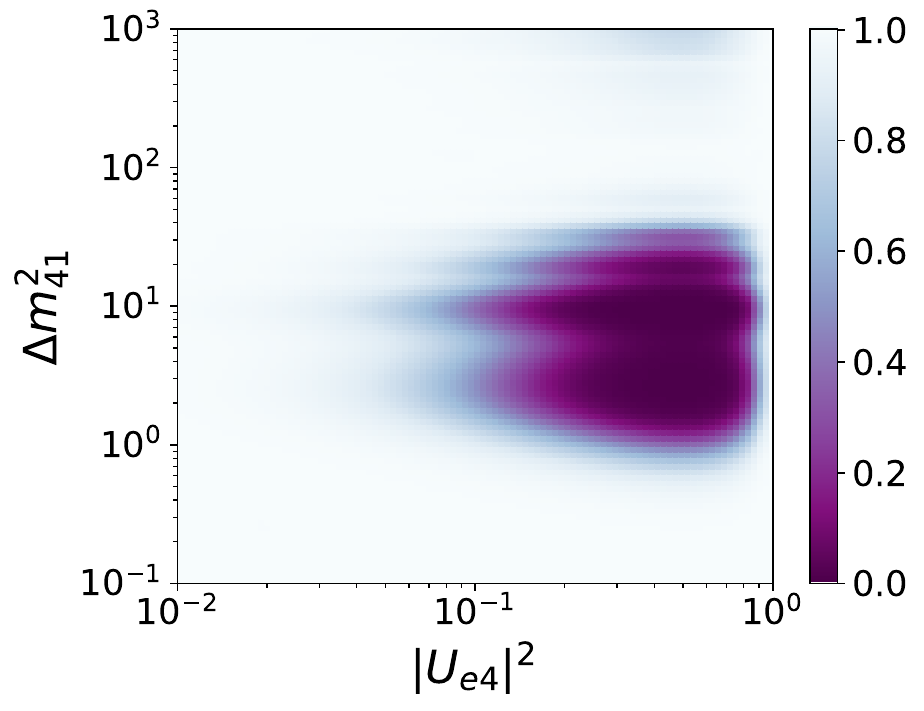}%
        \label{subfig:h}%
    }
    \caption{Same as Fig.~\ref{fig:analytical_1}, but for additional high-granularity binning choices.}
    \label{fig:analytical_2}
\end{figure*}

The resulting discrimination reach is shown in Figs.~\ref{fig:analytical_1} and \ref{fig:analytical_2} for several representative binning choices. In each panel, the entries in braces indicate the numbers of bins used in baseline, recoil energy, and timing, respectively. In this case, the discrimination power is given by $\mathbb{L}_{\rm NSI}$, which determines the color code in Figs.~\ref{fig:analytical_1} and \ref{fig:analytical_2}.
The plotted relative likelihood weights therefore visualize how strongly a sterile-generated dataset favors the sterile interpretation over the best-fitting NSI alternative across the benchmark parameter space. Equivalently, darker regions correspond to smaller values of $\mathbb{L}_{\rm NSI}$ and hence to stronger discrimination against the NSI hypothesis.

Several qualitative trends are immediately apparent. As expected, the fully inclusive \{1,1,1\} binning provides essentially no discriminatory power, since once all events are collapsed into a single counting bin, the sterile and NSI hypotheses can mimic one another modulo the level of the total event rate. In contrast, once multidimensional shape information is retained, the oscillatory structure characteristic of the sterile-neutrino scenario cannot, in general, be reproduced by the NSI benchmark, and substantial regions of parameter space become distinguishable.

The improvement is especially pronounced when the binning in baseline and recoil energy is refined. This behavior is physically well motivated: in the sterile-neutrino case, the signal is governed by an oscillation pattern that depends directly on $L/E_\nu$, so resolving the baseline and energy dependence is essential for exposing the characteristic departures from the NSI hypothesis. By comparison, increasing the number of timing bins yields a more modest improvement. Timing information remains useful, particularly because it partially separates the prompt and delayed source components, but in the benchmark setup studied here it is not as powerful a discriminator as the baseline and energy dependence.

\begin{figure}[t]
    \centering
    \includegraphics[width=0.5\linewidth]{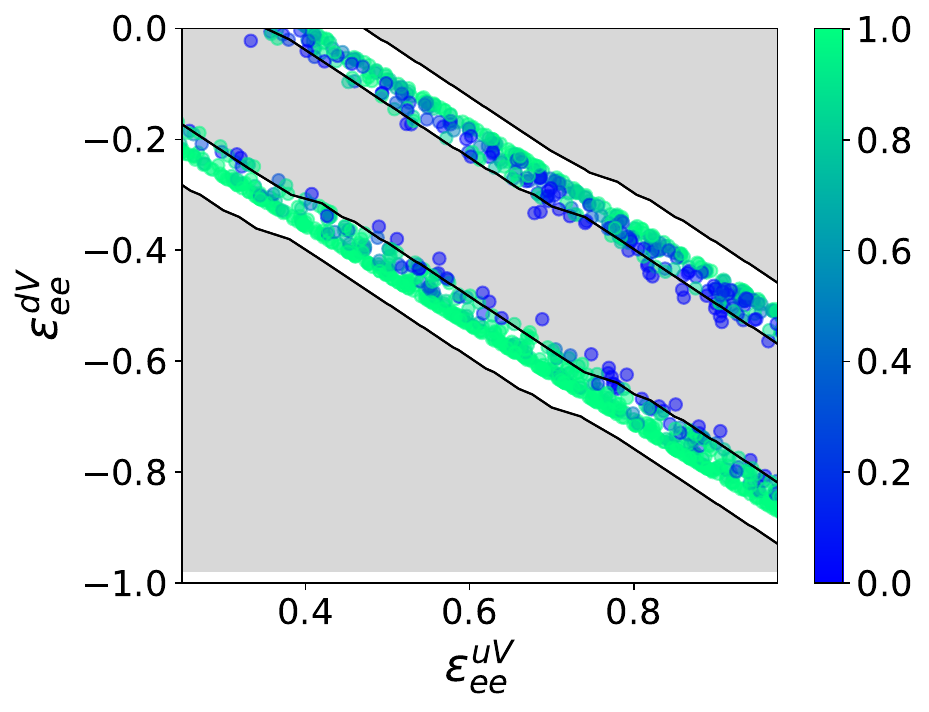}
    \caption{Mappings of best-fit NSI points from the sterile-neutrino model points sampled in Fig.~\ref{subfig:h}, colored according to their $\mathbb{L}_{\rm NSI}$ values compared against their corresponding sterile-neutrino datasets. The 99\% excluded regions from a parameter scan of a SM projected dataset are shown in grey for reference.}
    \label{fig:nsi-map}
\end{figure}

Figure~\ref{fig:nsi-map} shows a complementary view of the likelihood-based analysis by showing the points of the NSI parameter space that are selected as best-fits to their corresponding sterile-neutrino model points displayed in Fig.~\ref{subfig:h}. The points are colored according to their corresponding values of $\mathbb{L}_{\rm NSI}$. The excluded regions for an NSI parameter scan under a SM projected dataset are shown in grey at 99\% confidence, for reference. The excluded regions exhibit a characteristic double band structure, showing two allowed regions that lie around lines in the parameter space for which the resulting modifications in Eq. \ref{NSI_total_coupling} give similar rates to the SM. 
For sterile-neutrino model points with insignificant oscillation effects and very little rate deficit relative to SM, the best-fit NSI solutions tend to lie close to the central trajectory of one of the two allowed NSI bands. For these points, distinguishability is naturally challenging as very little indication of nonstandard effects are present in the data. By contrast, for sterile-neutrino model points with statistically more pronounced oscillatory structures, the best-fit NSI points are often displaced away from these central regions. This indicates that the parameter scans for those points are attempting to account for the total rate deficit imposed by the sterile-neutrino model via the NSI. The fact that many such points nevertheless yield very small values of $\mathbb{L}_{\rm NSI}$ make clear that distinguishability shown in Figs.~\ref{fig:analytical_1} and \ref{fig:analytical_2} is not driven solely by inclusive rate effects, but also by shape information that the NSI cannot faithfully account for. 

Taken together, these results show that even a conventional likelihood-based analysis can distinguish the two benchmark BSM scenarios in nontrivial regions of parameter space, provided that sufficient multidimensional information is retained. This establishes a useful baseline for the ML-based analyses presented in the following sections, where we examine whether more flexible inference methods can further enhance both model discrimination and parameter reconstruction.

\section{Neural Network Architecture and Training}
\label{Architecture}

The ML analyses that will be presented in Sec.~\ref{Binary} and Sec. \ref{Multi} are based on a common neural-network architecture and training procedure, which we summarize here. In our study, the binned CE$\nu$NS event distributions may be viewed as image-like objects: the events are organized in bins of baseline, reconstructed recoil energy, and timing, which play roles analogous to the two pixel dimensions and three color channels of an image. This representation makes convolutional neural networks (CNNs) a natural choice for the classification tasks considered here, in close analogy with their widespread use in neutrino-event classification and reconstruction~\citep{Aurisano:2016jvx,MicroBooNE:2016dpb,MicroBooNE:2018kka,Baldi:2018qhe,KM3NeT:2020zod,MicroBooNE:2020yze,DUNE:2020gpm,MicroBooNE:2020hho}.\footnote{Other network architectures (e.g., graph neural networks) could also be employed with appropriately processed data. However, comparing the performance of different network architectures is beyond the scope of this study.}

\begin{figure}[t]
    \centering
    \includegraphics[width=1.0\linewidth]{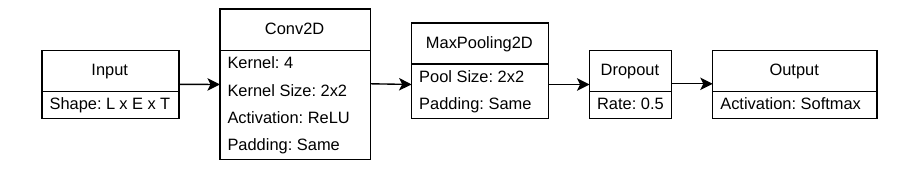}
    \caption{The machine architecture used in this study. Batch normalization operations are also included between layers. The shape of the input layer is determined by the total number of bins, as indicated, while the shape of the output layer depends on the number of sorting classes.}
    \label{fig:machine_arch}
\end{figure}

For a given binning choice $\{n_L,n_E,n_T\}$ in baseline, energy, and time, each pseudo-dataset is represented as an array of shape $n_L\times n_E\times n_T$. The baseline-energy plane is treated as a two-dimensional image, while the timing bins are interpreted as color channels. To process these inputs, we employ a shallow CNN implemented in the \texttt{Keras} deep-learning API. The network begins with an input layer matching the dimensions of the binned CE$\nu$NS tensor, followed by a two-dimensional convolutional layer with $4$ filters and a kernel size of $2\times2$, using the ReLU activation function. Batch-normalization layers are inserted between successive stages to stabilize training. The convolutional stage is followed by a two-dimensional max-pooling layer, a dropout layer with rate $0.5$ to curtail overtraining, and a final classification layer. For the binary-classification task in Sec. \ref{Binary}, the output layer contains two output nodes with softmax activation, corresponding to the sterile and NSI classes. For the multi-class task in Sec. \ref{Multi}, the number of output nodes is chosen to match the number of classes and a softmax activation is used. To ensure compatibility across different binning choices, padding (empty cells) is applied to the input when necessary so that the same network architecture can be used throughout the analysis. A schematic of the chosen architecture is shown in Fig.~\ref{fig:machine_arch}.

The training data are constructed from pseudo-datasets generated around the mean event predictions of the benchmark model points. Starting from the mean CE$\nu$NS distribution for a given point in parameter space, we generate large samples of fluctuated binned datasets, which are then labeled according to the underlying model hypothesis. In the current implementation, the fluctuations are applied independently to each bin using a Gaussian prescription. To focus the network on differences in multidimensional shape rather than on inclusive normalization, all pseudo-datasets are renormalized to a common total event rate before being passed to the classifier. The ML analysis is therefore intentionally insensitive to total-rate information and instead probes the discriminating power contained in the baseline-energy-timing structure of the CE$\nu$NS signal.
In particular, we emphasize that this is not intended as performance comparison relative to the likelihood-based analysis, but rather as a complementary method that is demonstrated in this different context.  The purpose is to emphasize what discrimination power is available to the neural network above and beyond standard approaches based on total rate.

Because the dominant flux-normalization uncertainty is largely removed by this normalization, we adopt a smaller fluctuation strength than in the likelihood-based analysis of Sec.~\ref{Analytical}. Specifically, we take $\sigma_{\eta}=0.05$ when adding  
gaussian fluctuations in creating the training data in the machine learning cases, whereas 
$\sigma_\eta=0.1$ was used in the likelihood-based analysis when integrating over the nuisance parameter $\eta$.

The resulting ensemble of labeled pseudo-datasets is then used for supervised training. 
The network is trained for $5$ epochs in Sec. \ref{Binary} and $10$ epochs in Sec. \ref{Multi}, with a validation split of $15\%$. We use \texttt{Keras}' categorical cross-entropy loss function and the Adam optimizer during training. After training, the network performance is evaluated on an independently generated test sample of fluctuated pseudo-datasets that is not used during training or validation. This is done to limit possible overfitting or bias. As mentioned earlier, the same architecture and training strategy are applied to the binary and multi-class classification problems described in Sec.~\ref{Binary} and Sec.~\ref{Multi}, respectively.

\section{Machine-Learning-Based Model Discrimination} \label{sec:MLdisc}

We now present the results of the ML-based analyses introduced in the previous section. As discussed above, the binned CE$\nu$NS event distributions are treated as image-like inputs to a CNN, allowing the classifier to learn discriminating features directly from the multidimensional baseline-energy-timing structure of the signal. We consider two complementary tasks: binary classification between the sterile-neutrino and best-fit NSI hypotheses, and multi-class classification aimed at resolving different regions of the underlying parameter space.

\subsection{Machine-learning-based binary classification}
\label{Binary}

We begin with the binary-classification task, in which the network is trained to distinguish sterile-generated CE$\nu$NS datasets from those generated under the corresponding best-fit NSI hypothesis. For each point in the scanned sterile-neutrino parameter space, an independent classifier is trained and tested, allowing the classification performance to be mapped directly onto the benchmark parameter space. In this way, the binary-classification analysis provides a measure of how readily the sterile and NSI scenarios can be separated using the shape information contained in the binned event distributions alone.

\begin{figure*}[t]
\centering
    \subfloat[\{1,1,1\}]{%
        \includegraphics[width=.48\linewidth]{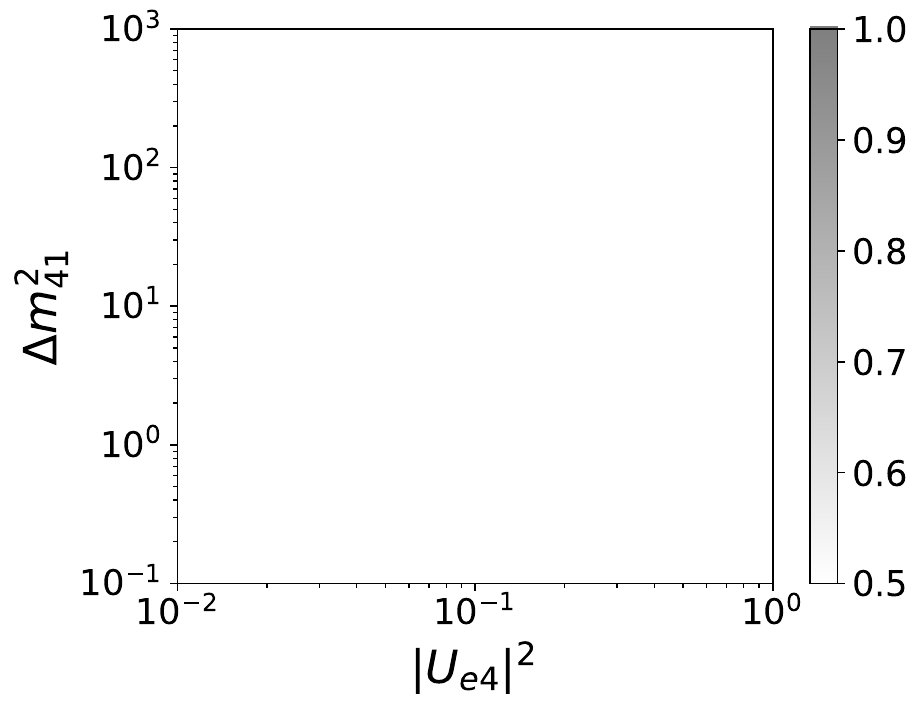}%
        \label{subfig:a}%
    }
    \subfloat[\{1,9,3\}]{%
        \includegraphics[width=.48\linewidth]{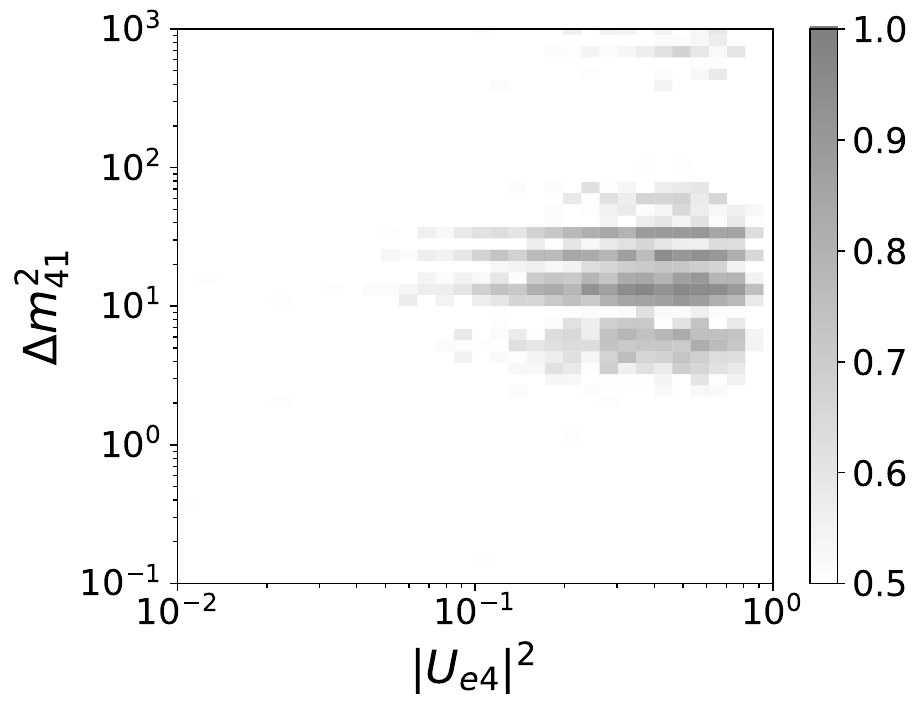}%
        \label{subfig:b}%
    }\\
    \subfloat[\{3,3,1\}]{%
        \includegraphics[width=.48\linewidth]{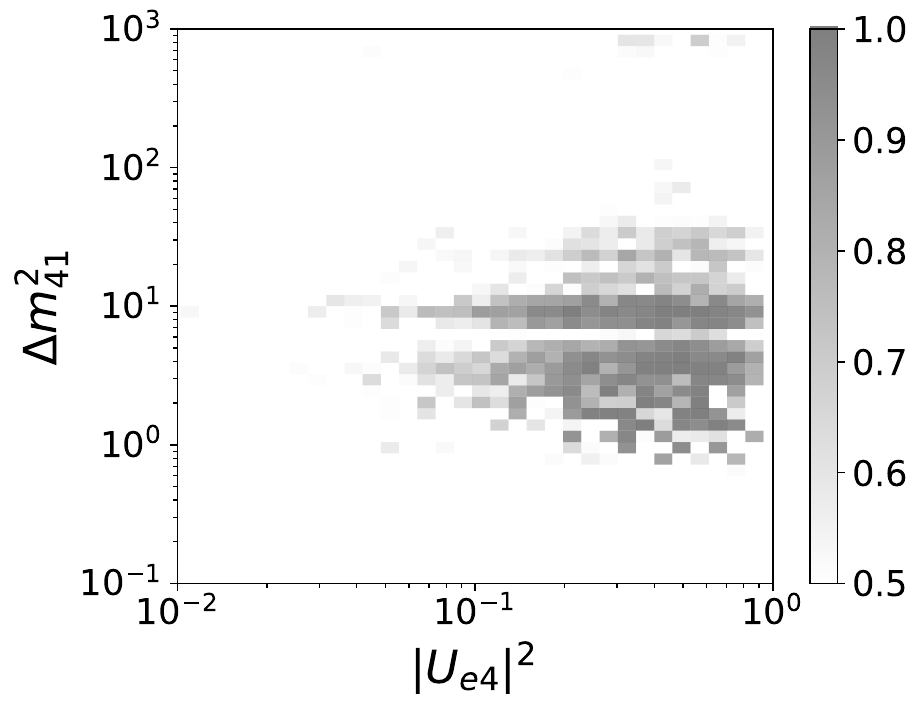}%
        \label{subfig:c}%
    }
    \subfloat[\{3,9,3\}]{%
        \includegraphics[width=.48\linewidth]{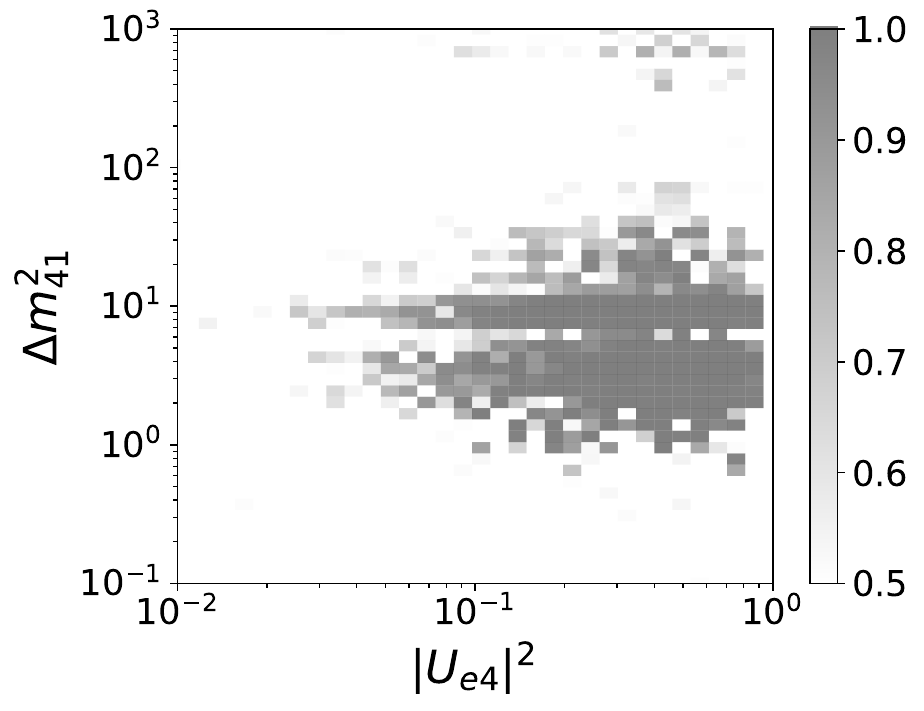}%
        \label{subfig:d}%
    }
    \caption{Classification accuracy for the binary classification between NSI data and data from points of the $\nu_{e}$-coupled sterile model space.
        Several binning configurations are shown, indicating the number of bins in each dimension as \{Length, Energy, Time\}. The accuracy is shown ranging from 0.5 (random guessing) to 1.0 (perfect accuracy).}
    \label{fig:binary_1}
\end{figure*}

\begin{figure*}[t]
\centering
    \subfloat[\{9,1,3\}]{%
        \includegraphics[width=.48\linewidth]{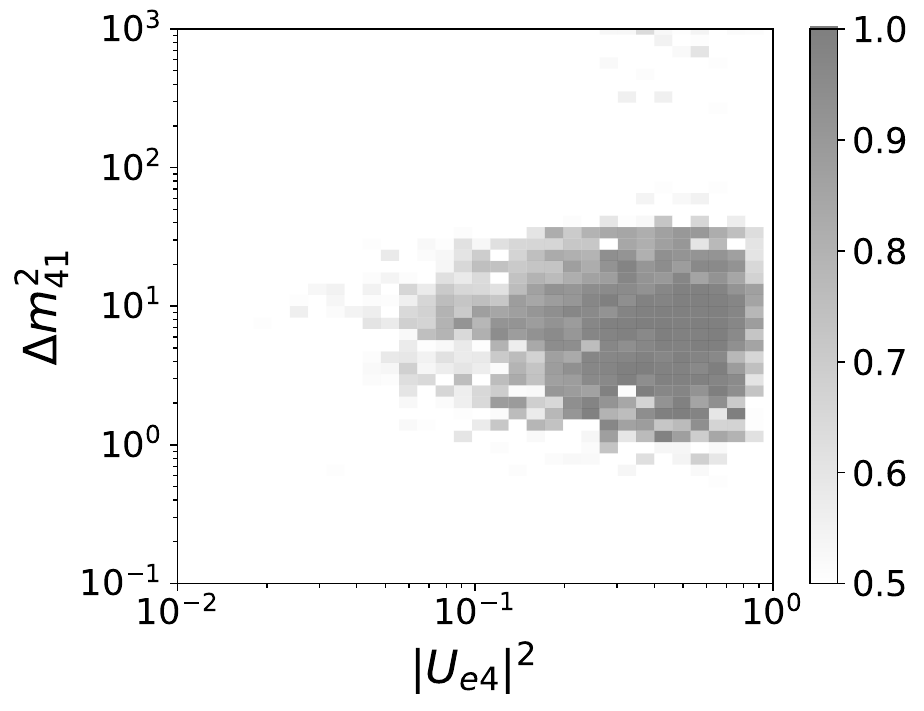}%
        \label{subfig:a}%
    }
    \subfloat[\{9,3,3\}]{%
        \includegraphics[width=.48\linewidth]{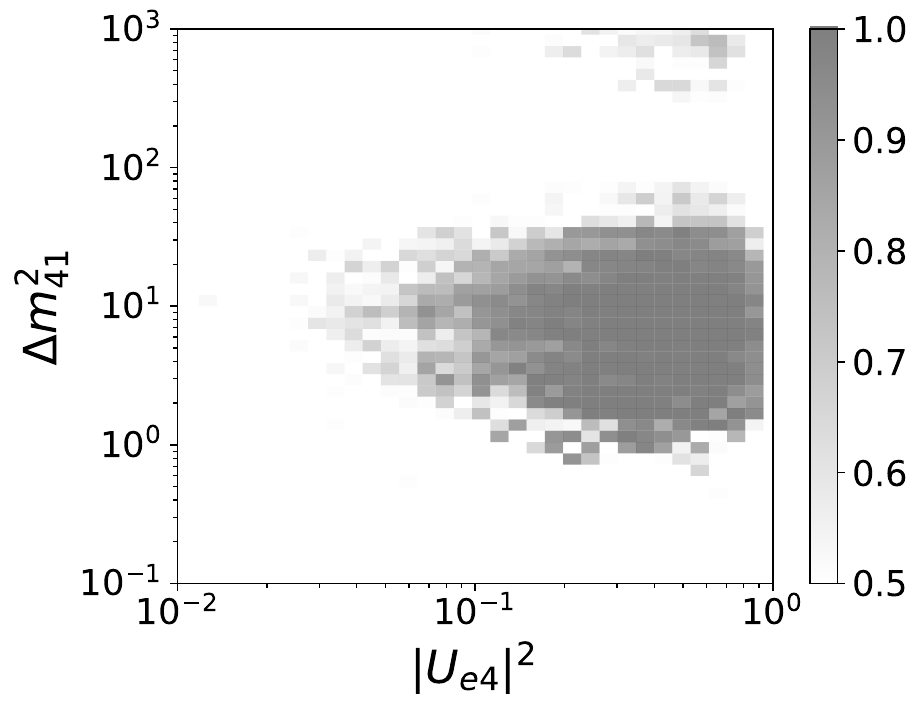}%
        \label{subfig:b}%
    }\\
    \subfloat[\{9,9,1\}]{%
        \includegraphics[width=.48\linewidth]{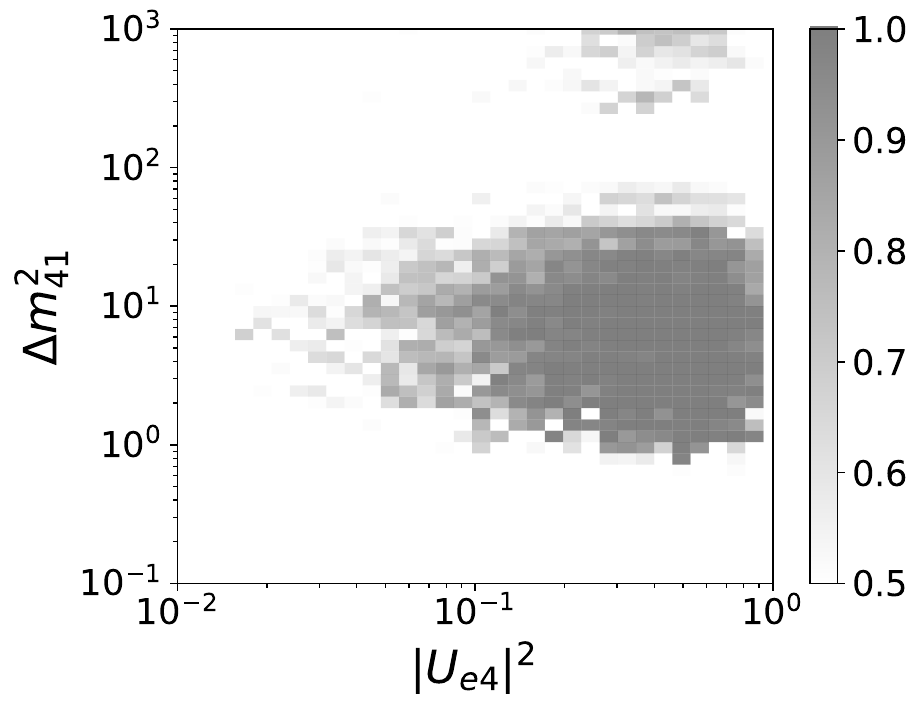}%
        \label{subfig:c}%
    }
    \subfloat[\{9,9,3\}]{%
        \includegraphics[width=.48\linewidth]{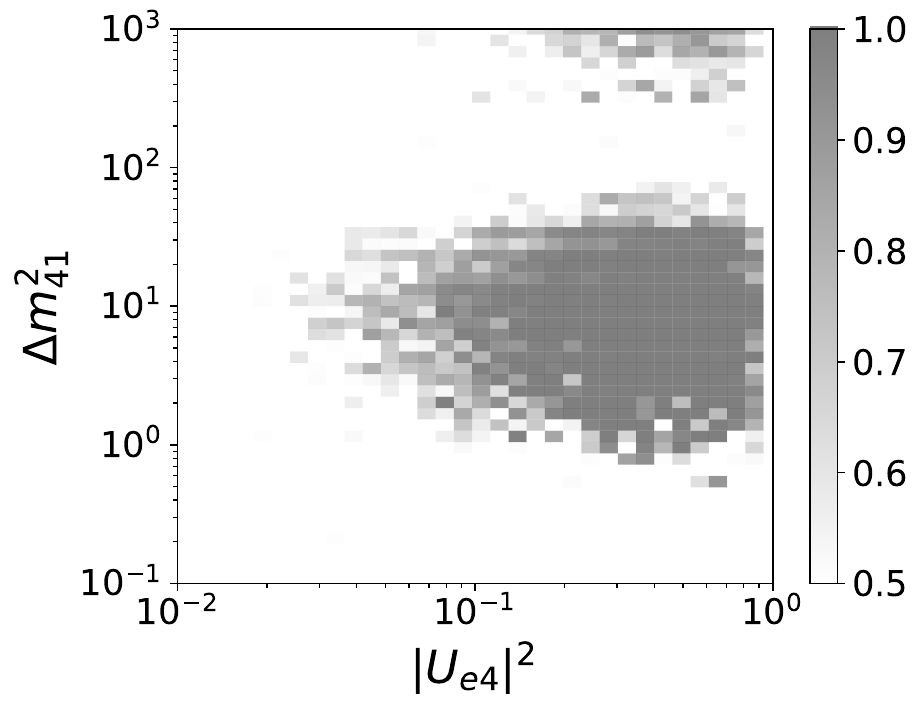}%
        \label{subfig:d}%
    }
    \caption{Same as Fig.~\ref{fig:binary_1} for additional binning choices.}
    \label{fig:binary_2}
\end{figure*}

For each point in the scanned sterile-neutrino parameter space, we first identify the corresponding best-fit NSI point again using \texttt{MultiNest}~\citep{Feroz:2007kg, Feroz:2008xx, Feroz:2013hea}. Following the pseudo-data generation procedure described in Sec.~\ref{Architecture}, we then construct training samples for the binary task from $10^4$ fluctuated datasets generated under the sterile benchmark point and an equal number generated under its best-fit NSI counterpart, yielding a balanced set of training samples.
A separate neural network is trained for each sterile benchmark point. The trained network is evaluated on an independently generated held-out test sample consisting of equal numbers of sterile-generated and NSI-generated pseudo-datasets. The classification accuracy is then defined in the standard way as the fraction of correctly classified examples in this balanced binary test set,
\begin{equation}
\mathrm{Accuracy}
=
\frac{N_{\rm correct}}{N_{\rm test}}
=
\frac{N_{\rm sterile \to sterile}+N_{\rm NSI \to NSI}}
{N_{\rm sterile}+N_{\rm NSI}},
\end{equation}
where $N_{\rm sterile \to sterile}$ and $N_{\rm NSI \to NSI}$ denote the numbers of correctly classified sterile and NSI test samples, respectively. In the present setup, the held-out test sample contains $100$ pseudo-datasets from each class, so that $N_{\rm test}=200$. An accuracy of $0.5$ therefore corresponds to no effective discrimination on the balanced test set (i.e., random guessing), while an accuracy of $1.0$ corresponds to perfect separation between the sterile and NSI hypotheses. Higher accuracy therefore indicates that the trained network more reliably distinguishes sterile-generated datasets from those generated under the best-fit NSI alternative, (or vice versa). The resulting classification accuracies across the sterile parameter space are shown in Figs.~\ref{fig:binary_1} and \ref{fig:binary_2} for several representative binning choices, labeled by the numbers of bins in \{baseline, energy, time\}. The classification accuracy shown by the gray color scale ranges from $0.5$ to $1.0$.

Overall, the binary-classification maps exhibit qualitative trends similar to those found in the likelihood-based analysis, providing a useful cross-check that the network is responding to the same underlying physical structure of the signal. More importantly, however, the present results demonstrate that substantial model-discrimination power remains even after the total event rate has been explicitly removed from the input. The CNN is therefore not relying on inclusive normalization differences, but is instead learning features associated with the multidimensional shape of the CE$\nu$NS distribution.

This point is particularly significant in the present context. In the sterile-neutrino benchmark, the dominant departures from the NSI hypothesis arise through correlated distortions in the baseline and recoil-energy dependence of the signal, with timing information playing a secondary but still useful role. The fact that the classifier achieves high accuracy over substantial regions of parameter space indicates that these oscillation-induced shape features are sufficiently distinctive to support robust model separation on their own. From this perspective, the ML-based binary analysis should be viewed not merely as a repetition of the likelihood-based study, but as evidence that the discriminatory information is genuinely encoded in the shape of the data. This is encouraging for applications in which normalization uncertainties are large or challenging to control (for example, at reactor neutrino sources), since it suggests that rate-insensitive inference strategies may still provide meaningful separation between competing new-physics scenarios.

One may also consider the reverse construction, namely generating pseudo-datasets from selected points in the NSI parameter space, determining the corresponding best-fit sterile-neutrino points, and then training the network to discriminate between fluctuations of each NSI benchmark and its sterile counterpart.
We do not pursue that analysis in this study. In the restricted benchmark setup adopted here, the reverse exercise is expected to be less informative than the sterile-generated study shown in Figs.~\ref{fig:binary_1} and \ref{fig:binary_2}. This is because for many of the NSI points considered here, the dominant observable effect is a change in the effective $\nu_e$ scattering rate, which can often be approximately mimicked by a sterile-neutrino hypothesis if the oscillation pattern is sufficiently washed out over the experimentally accessible range. In such cases, the reverse scan would be dominated by regions with limited shape-based discrimination, while requiring a substantially larger computational cost. For this reason, we focus on the sterile-generated construction, where the physically distinctive oscillation structure provides the more informative benchmark for assessing model discernment. This should be understood as a practical limitation of the present benchmark study, rather than a general statement about arbitrary sterile-neutrino and NSI scenarios.

\subsection{Machine multi-classification}
\label{Multi}

We next consider a more demanding neural-network task, namely multi-class classification within the sterile-neutrino parameter space itself. As in the binary-classification study, pseudo-datasets are generated from benchmark points in the chosen sterile parameter window following the procedure described in Sec.~\ref{Architecture}. The essential difference is that, rather than training the network to distinguish the sterile hypothesis from the best-fit NSI alternative, we now train a single network to distinguish among multiple cells within the sterile parameter space. In this way, the analysis moves beyond model discrimination and addresses a more refined question: whether the multidimensional CE$\nu$NS event distribution contains enough information for the network to identify the approximate location of the underlying sterile-neutrino parameter point.

To carry out the multi-classification study, we divide the chosen sterile-neutrino parameter window into a $20\times 20$ grid and take the center of each cell as a representative benchmark point. For every cell, we generate $10^4$ fluctuated pseudo-datasets for training, and after training, we evaluate the network on an independently generated test sample containing $10^3$ pseudo-datasets for that same cell.

In contrast to the binary-classification task discussed in Sec.~\ref{Binary}, the purpose here is not simply to decide between two competing hypotheses, but rather to determine how well the network can identify the approximate location of the underlying sterile-neutrino parameter point. Accordingly, the performance measure is based on the average distance between the true cell and the cell selected by the network, computed separately along the horizontal axis ($|U_{e4}|^2$) and the vertical axis ($\Delta m_{41}^2$).

\begin{figure*}[t]
\centering
    \subfloat[\{1,1,1\}]{%
        \includegraphics[width=.48\linewidth]{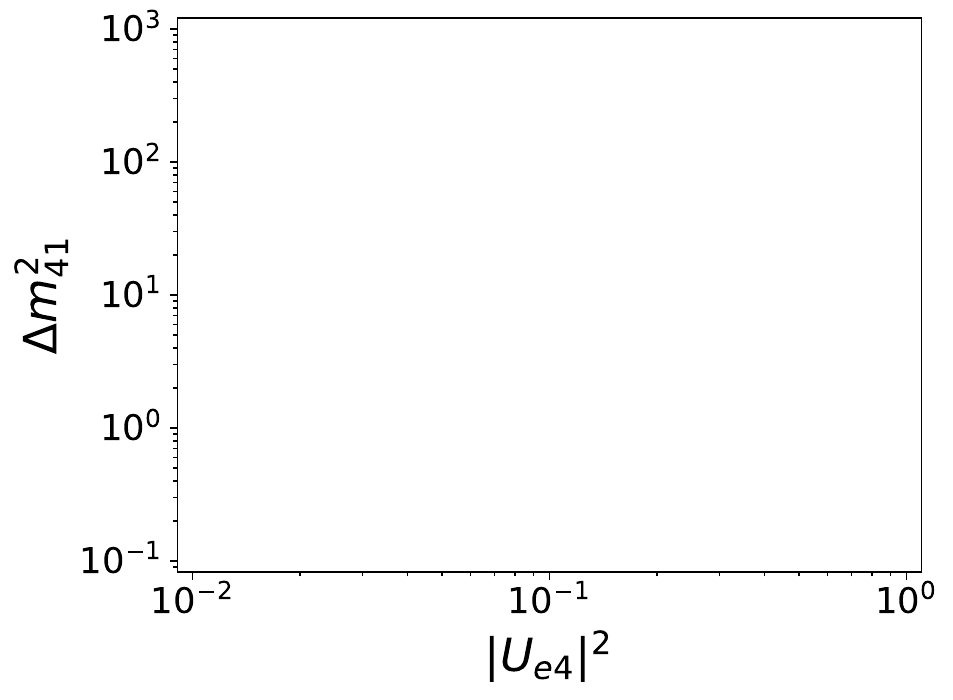}%
        \label{subfig:a}%
    }
    \subfloat[\{1,9,3\}]{%
        \includegraphics[width=.48\linewidth]{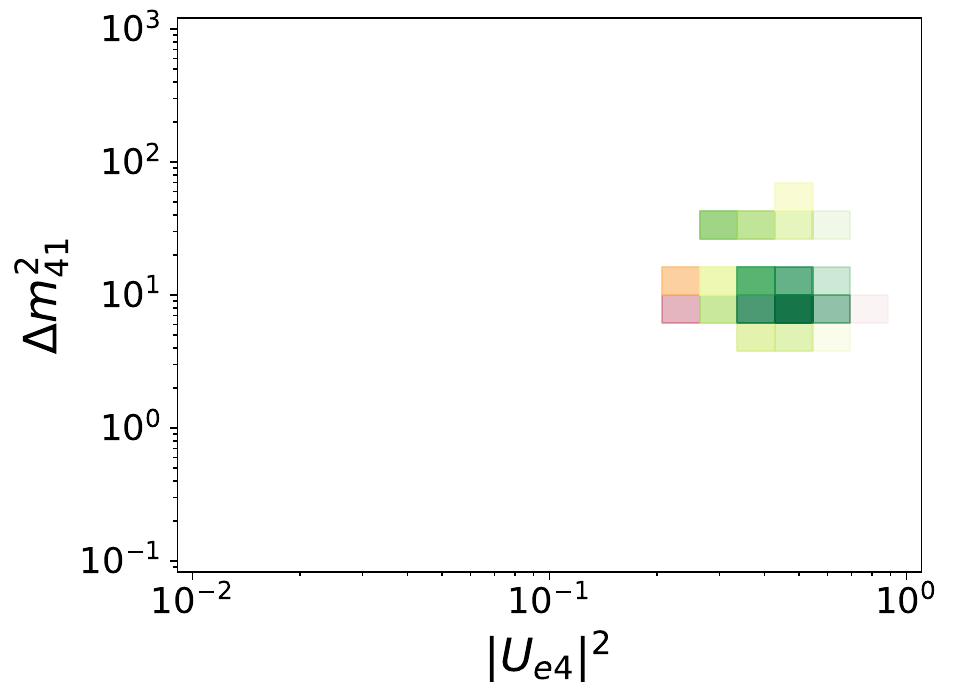}%
        \label{subfig:b}%
    }\\
    \subfloat[\{3,3,1\}]{%
        \includegraphics[width=.48\linewidth]{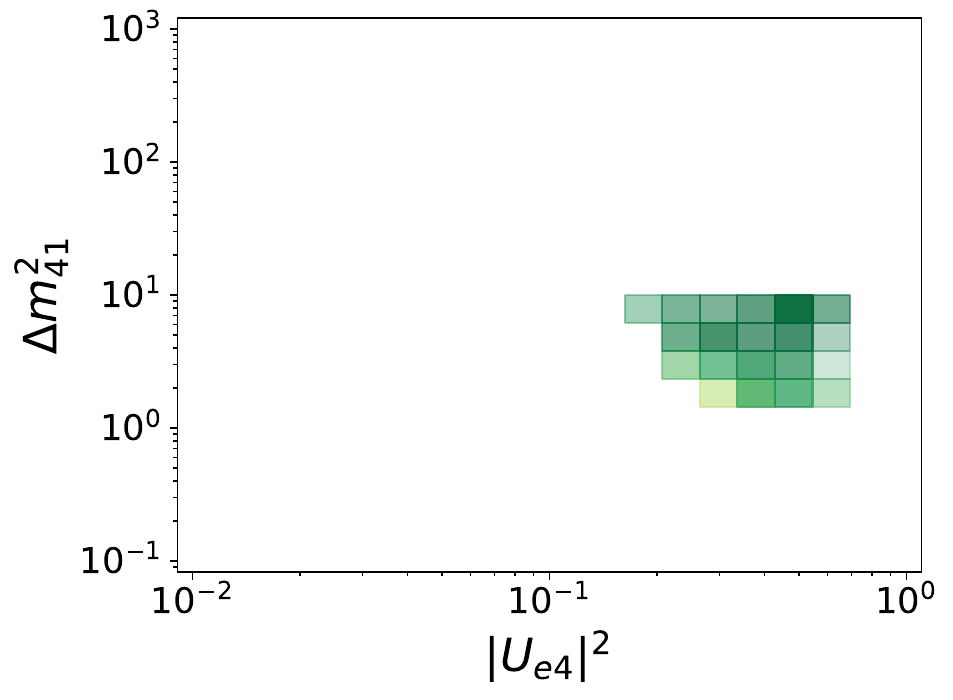}%
        \label{subfig:c}%
    }
    \subfloat[\{3,9,3\}]{%
        \includegraphics[width=.48\linewidth]{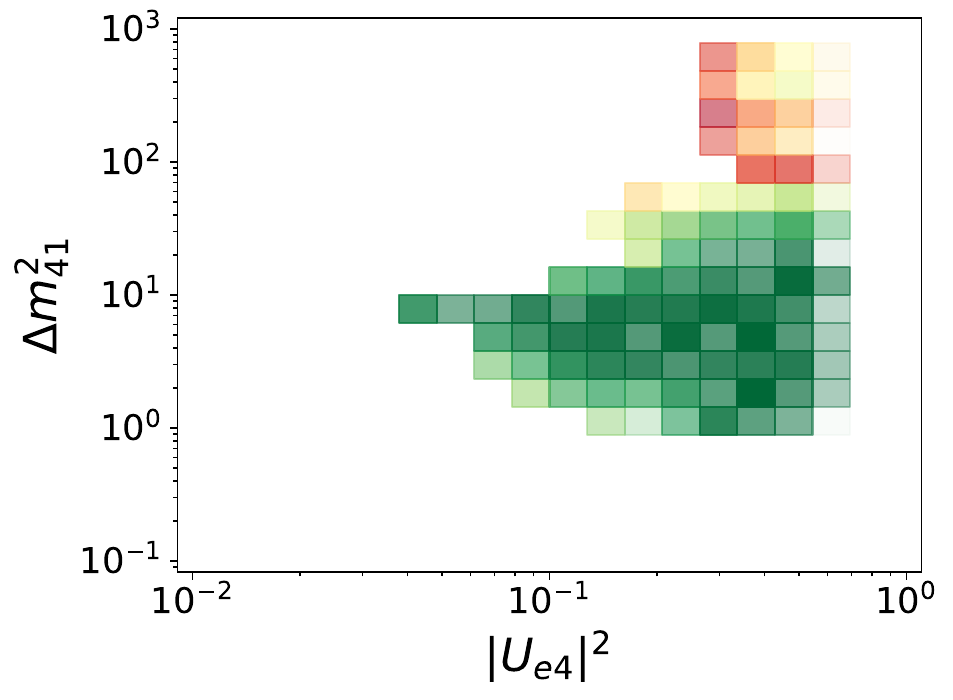}%
        \label{subfig:d}%
    }
    
    \caption{
Multi-class localization accuracy in the $\nu_e$-coupled sterile-neutrino parameter space, with opacity indicating horizontal error in $|U_{e4}|^2$ and color indicating vertical error in $\Delta m^2_{41}$ (shown for cells with average activation $> 0.05$).
Several binning configurations are shown, indicating the number of bins in each dimension as \{Length, Energy, Time\}. }
    \label{fig:multi_1}
\end{figure*}

\begin{figure*}[t]
\centering
    \subfloat[\{9,1,3\}]{%
        \includegraphics[width=.48\linewidth]{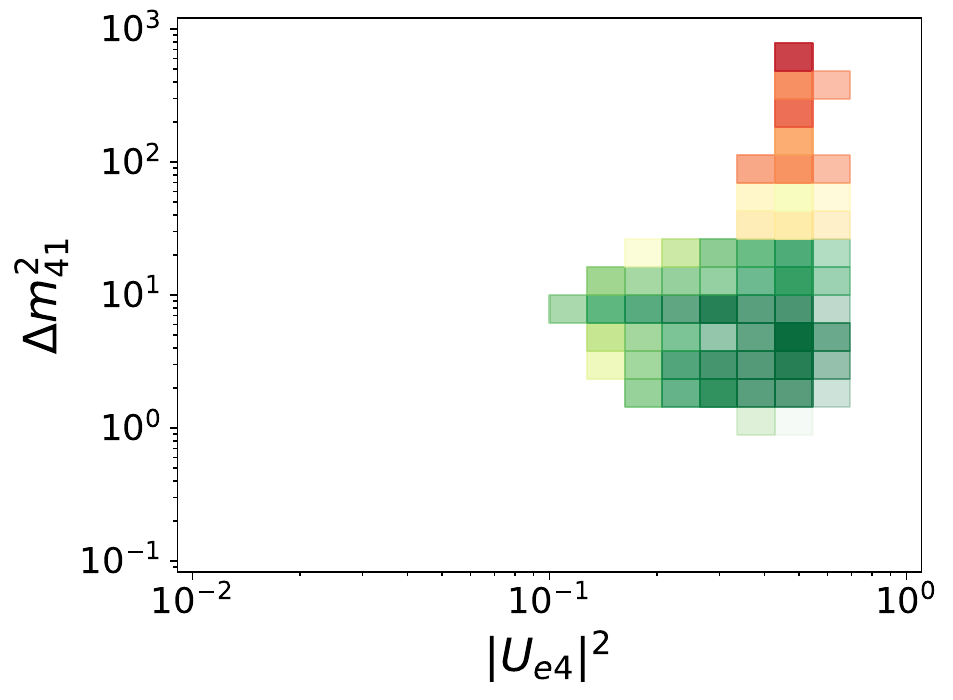}%
        \label{subfig:a}%
    }
    \subfloat[\{9,3,3\}]{%
        \includegraphics[width=.48\linewidth]{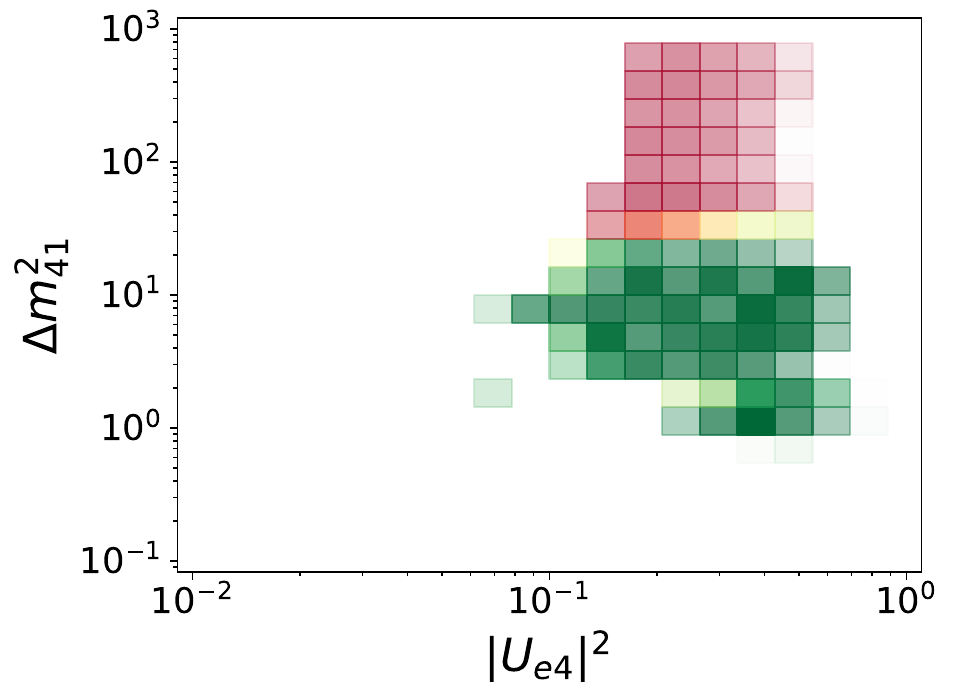}%
        \label{subfig:b}%
    }\\
    \subfloat[\{9,9,1\}]{%
        \includegraphics[width=.48\linewidth]{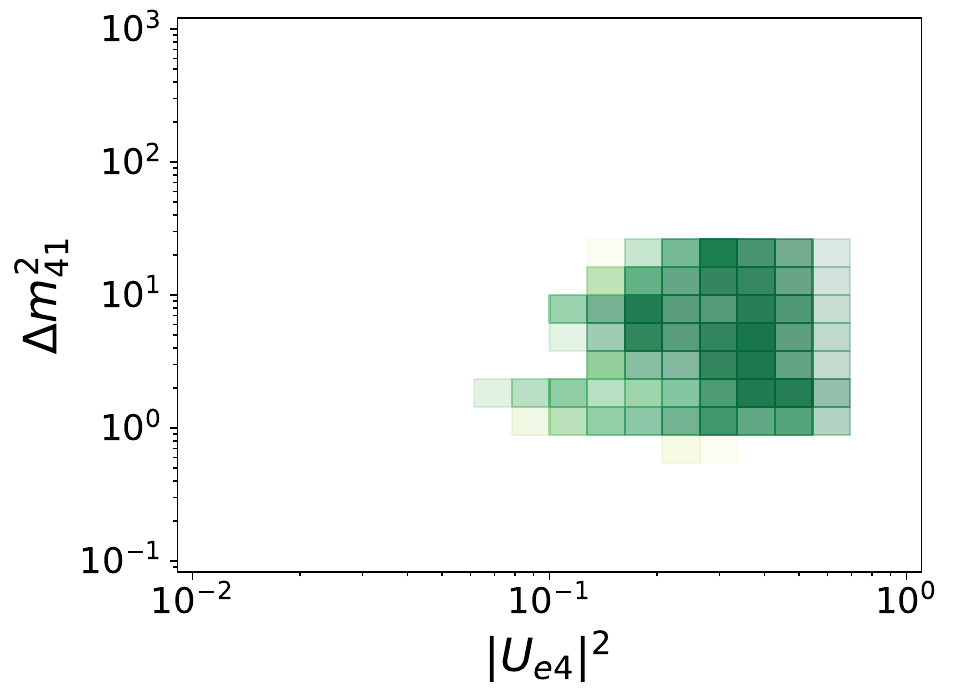}%
        \label{subfig:c}%
    }
    \subfloat[\{9,9,3\}]{%
        \includegraphics[width=.48\linewidth]{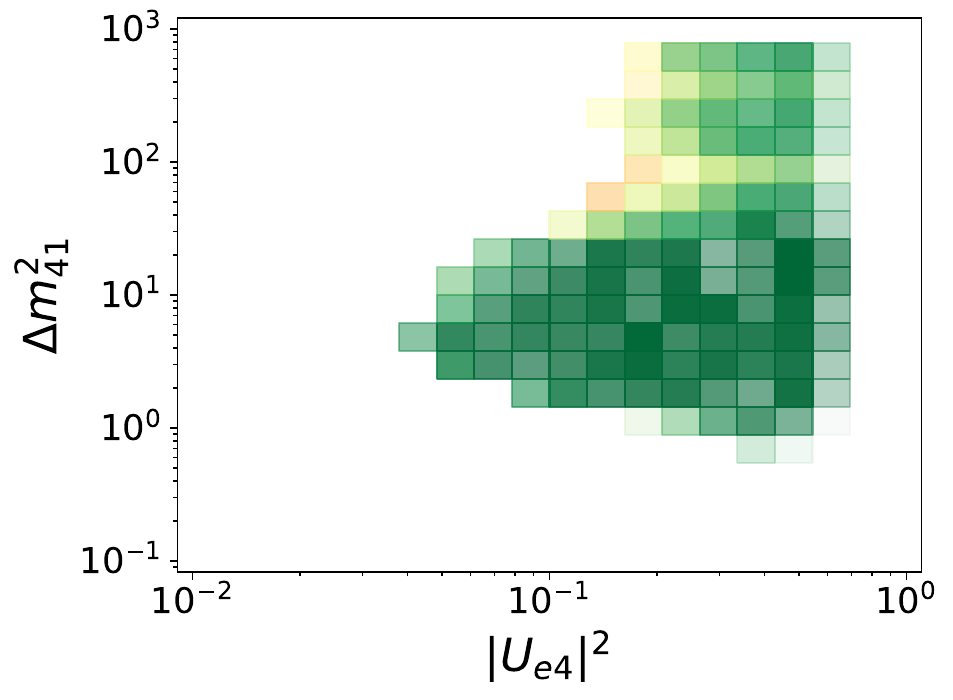}%
        \label{subfig:d}%
    }
    \caption{Same as Fig.~\ref{fig:multi_1} for additional binning choices.}
    \label{fig:multi_2}
\end{figure*}

To display the multi-classification results in a compact and interpretable way, we retain only those cells for which the average node activation of the network’s preferred class exceeds a minimum threshold. In practice, this threshold is introduced as a visualization criterion to suppress low-confidence assignments that would otherwise populate the map with noisy or weakly informative cells. We choose a value of $0.05$, which was found empirically to provide a reasonable balance between retaining coverage of the parameter space and highlighting regions where the classifier exhibits meaningful localization performance.\footnote{The qualitative features of the multi-classification results remain stable under modest variations of this visualization threshold around the chosen value.} Each surviving cell is then assigned an opacity and a color that summarize the localization performance. The opacity encodes the average horizontal displacement: full opacity corresponds to perfect reconstruction in $|U_{e4}|^2$, while zero opacity indicates an average error of three cells or more. The color similarly encodes the average vertical displacement in $\Delta m_{41}^2$, with green indicating perfect reconstruction and red indicating an average error of three cells or more.

Results of this analysis for a suite of different binning choices are shown in Figs.~\ref{fig:multi_1} and \ref{fig:multi_2}. Several qualitative features are immediately apparent. First, the fully inclusive \{1,1,1\} binning provides essentially no meaningful parameter-space localization, indicating that once the CE$\nu$NS signal is reduced to a single counting bin, the network cannot reliably infer the underlying sterile-neutrino benchmark point.

Second, once baseline and recoil-energy binning are introduced, nontrivial regions of the sterile parameter space begin to show useful localization performance. In particular, the opacity patterns indicate that the network can recover the horizontal direction, corresponding to $|U_{e4}|^2$, with reasonable fidelity over substantial regions of parameter space, while the cell colors show that the vertical direction, corresponding to $\Delta m_{41}^2$, is in general more difficult to reconstruct. This behavior is physically sensible. The mixing parameter primarily controls the overall strength of the oscillation-induced distortion, whereas the mass-squared splitting governs the oscillation scale itself; the latter is more easily degraded by recoil-energy smearing, finite binning, and partial averaging of the oscillation pattern.

Third, the most successful reconstruction is obtained in intermediate and relatively large-mixing regions, where the sterile-induced distortions are strong enough to imprint a recognizable multidimensional pattern on the CE$\nu$NS distribution. By contrast, in regions with smaller mixing or in parts of the large-$\Delta m_{41}^2$ regime, the network often fails to achieve confident localization, as reflected either by reduced opacity or by the absence of plotted cells after the activation threshold is applied. In this sense, the blank or weakly shaded regions should be interpreted not merely as poorer classification performance, but as regions where the available observables do not support robust parameter-space identification with the present setup.

Finally, the overall trend with binning again highlights the importance of multidimensional shape information. Increasing the granularity in baseline and recoil energy leads to the most significant gains, while timing information provides an additional but subleading improvement. The multi-classification results therefore reinforce the broader conclusion of this work: the stopped-pion CE$\nu$NS signal contains sufficient structured information not only to discriminate between benchmark BSM scenarios, but also, in favorable regions of parameter space, to localize the underlying sterile-neutrino parameters themselves.

\section{Discussion and Conclusion} \label{sec:conlcusions}

In this work, we have investigated how stopped-pion CE$\nu$NS experiments can be used not only to detect deviations from the standard three-neutrino expectation, but also to discriminate among candidate new-physics explanations and to infer the underlying parameter region responsible for an observed signal. Using the $3+1$ sterile-neutrino scenario and neutral-current NSI as benchmark frameworks, we have shown that the correlated baseline, recoil-energy, and timing structure of the CE$\nu$NS signal provides nontrivial discriminatory power beyond what can be obtained from inclusive event rates alone.

At the level of model discrimination, our likelihood-based analysis established a useful baseline by quantifying the regions of sterile parameter space for which sterile-generated datasets can be distinguished from their best-fit NSI counterparts. As expected, discrimination is weak when the data are reduced to a single counting bin, but becomes substantially stronger once multidimensional shape information is retained. In particular, the baseline and recoil-energy structure of the signal play the dominant role, while timing provides an additional but comparatively more modest handle.

The ML analyses reinforce and extend this picture. In the binary classification study, we found that a CNN trained on rate-renormalized pseudo-datasets is still able to separate the sterile and NSI benchmarks over substantial regions of parameter space. This is an important observation, because it shows that the distinguishing information is not merely a consequence of normalization differences, but is genuinely encoded in the multidimensional shape distortions of the CE$\nu$NS event distribution. In this sense, the ML approach should be viewed not simply as a cross-check of the likelihood analysis, but as evidence that rate-insensitive inference strategies can retain significant discriminatory power even in the presence of large normalization uncertainties.

The multi-classification analysis further demonstrates that, in favorable regions of sterile parameter space, the CE$\nu$NS signal contains enough structured information not only for model discrimination but also for approximate parameter-space localization. Although this reconstruction is necessarily coarser and more challenging than binary separation, the results show that the same baseline-energy-timing correlations that support model discernment can also be used to identify the approximate location of the underlying sterile benchmark point. The reconstruction is generally more robust in the $|U_{e4}|^2$ direction than in the $\Delta m_{41}^2$ direction, which is physically consistent with the differing ways in which these parameters control the strength and scale of the oscillation pattern.

Taken together, these results provide a proof of concept that ongoing CE$\nu$NS experiments with multiple detectors at stopped-pion sources offer a promising arena not only for testing the presence of nonstandard neutrino-sector physics, but also for interpreting its origin. More broadly, the methods developed here illustrate how conventional likelihood-based analyses and ML-based inference can play complementary roles in moving from anomaly detection to physics interpretation. Such strategies may become increasingly valuable in future neutrino and dark-sector searches, where low statistics, sizable systematic uncertainties, and partial degeneracies among candidate models can otherwise obscure the underlying physics.

There are several natural directions for future work. On the phenomenological side, the benchmark study considered here can be extended to broader parameter spaces and to additional classes of new-physics scenarios, including light dark matter signals that may mimic or compete with sterile-neutrino and NSI interpretations. Within the restricted benchmark setup adopted here, we have focused on sterile-generated pseudo-datasets, since the reverse construction based on NSI-generated samples is expected to be less informative while incurring substantially greater computational cost; a more symmetric treatment of both directions would nevertheless be worthwhile in broader future studies. On the methodological side, it will be important to incorporate more realistic detector effects, systematic-uncertainty treatments, and alternative network architectures, as well as to explore direct parameter-regression strategies beyond the classification-based framework adopted here. Finally, because the present analysis makes clear which observables most strongly enhance model discernment, these methods may also help inform the design of the ongoing and proposed CE$\nu$NS detector configurations 
by clarifying the detector capabilities and source characteristics needed to achieve robust discrimination between competing new-physics hypotheses.

\acknowledgments
The work of IAB and BD is supported by the U.S. Department of Energy Grant DE-SC0010813. The work of JWW is supported by the National Science Foundation grant PHY-2412995.

\bibliographystyle{JHEP}
\bibliography{refs.bib}

\end{document}